\documentclass[11pt]{article}

\usepackage{float}

\usepackage{tikz}
\usetikzlibrary{shapes,arrows} 

\usepackage[label font=it]{subfig}
\usepackage{caption}
\usepackage{multirow,tabularx}

\newsavebox{\twofigures}
\usepackage[export]{adjustbox}
\usepackage{calc}
\usepackage[version=4]{mhchem}
\usepackage{siunitx}
\usepackage{longtable,tabularx}
\usepackage{tabularray}
\usepackage{lmodern,pdftexcmds,amsmath}

\usepackage[title]{appendix}
\usepackage{titlesec}
\titleformat{\paragraph}
{\normalfont\normalsize}{\theparagraph}{1em}{}
\titlespacing*{\paragraph}
{0pt}{3.25ex plus 1ex minus .2ex}{1.5ex plus .2ex}
\counterwithin*{equation}{section}
\DeclareMathAlphabet{\mathsfbi}{OT1}{\sfdefault}{bx}{sl}
\DeclareMathVersion{sfletters}
\SetSymbolFont{letters}{sfletters}{OML}{ntxsfmi}{b}{it}

\makeatletter
\newcommand{\mathbfsbilow}[1]{%
  \text{\mathversion{sfletters}$\m@th#1$}%
}

\DeclareRobustCommand{\tensor}[1]{%
  \begingroup
  \ifcat\noexpand #1\relax
    \edef\greek@test{\detokenize{#1}}%
    \edef\greek@test{\expandafter\@cdr\greek@test\@nil}%
    \edef\greek@test{\expandafter\@car\greek@test\@nil}%
    \edef\x{\the\lccode\expandafter`\greek@test}%
    \edef\y{\number\expandafter`\greek@test}%
    \ifnum\x=\y\relax
      \mathbfsbilow{#1}%
    \else
      \mathsfbi{#1}%
    \fi
  \else
    \mathsfbi{#1}%
  \fi
  \endgroup
}
\makeatother

\usepackage{orcidlink}

\usepackage{tabularx, caption}
 \makeatletter
 \newcommand*{\compress}{\@minipagetrue}
 \makeatother

\newsavebox{\bigimage}

\usepackage{graphicx}
\usepackage{newtxtext}
\usepackage{newtxmath}
\usepackage{natbib}
\usepackage{hyperref}

\usepackage{algorithm}
\usepackage[noend]{algpseudocode}

\usepackage{etoolbox} 
\makeatletter
\patchcmd{\NAT@citex}
  {\@citea\NAT@hyper@{%
     \NAT@nmfmt{\NAT@nm}%
     \hyper@natlinkbreak{\NAT@aysep\NAT@spacechar}{\@citeb\@extra@b@citeb}%
     \NAT@date}}
  {\@citea\NAT@nmfmt{\NAT@nm}%
   \NAT@aysep\NAT@spacechar\NAT@hyper@{\NAT@date}}{}{}

\patchcmd{\NAT@citex}
  {\@citea\NAT@hyper@{%
     \NAT@nmfmt{\NAT@nm}%
     \hyper@natlinkbreak{\NAT@spacechar\NAT@@open\if*#1*\else#1\NAT@spacechar\fi}%
       {\@citeb\@extra@b@citeb}%
     \NAT@date}}
  {\@citea\NAT@nmfmt{\NAT@nm}%
   \NAT@spacechar\NAT@@open\if*#1*\else#1\NAT@spacechar\fi\NAT@hyper@{\NAT@date}}
  {}{}
\makeatother

\hypersetup{
    colorlinks = true,
     citecolor  = blue,
    allcolors=blue,
}

\newcommand{\RomanNumeralCaps}[1]

\usepackage{multirow,tabularx}

\usepackage[export]{adjustbox}
\usepackage{calc}
\usepackage[version=4]{mhchem}
\usepackage{siunitx}
\usepackage{longtable,tabularx}
\usepackage{tabularray}
 \UseTblrLibrary{amsmath}
 \usepackage{lmodern,pdftexcmds,amsmath}

\usepackage[noblocks]{authblk}
\usepackage[margin=1in]{geometry}
\usepackage[title]{appendix}

\newcommand\affiliation[1]{\gdef\@affiliation{\let\aff\aff@inst#1}}
\gdef\@affiliation{}
\def\email#1{Email address for correspondence: #1}
\def\aff#1{\ignorespaces\textsuperscript{#1}}
\def\corresp#1{\unskip\thanks{#1}}
\numberwithin{equation}{section}

\renewenvironment{abstract}
{\begin{quote}
\noindent \rule{\linewidth}{.5pt}\par{\bfseries \abstractname.}}
{\medskip\noindent \rule{\linewidth}{.5pt}
\end{quote}
}

\hypersetup{
   colorlinks=true,
   linkcolor={blue!100!black},
   citecolor={blue!100!black},
   plainpages=false,
}

\title{\bf Resolvent-based estimation of a turbulent wake}

\author[1, 2]{\bf Junoh Jung\corresp{\email{junohj@umich.edu}}}
\author[1]{\bf Aaron Towne}

\affil[1]{\normalsize Department of Mechanical Engineering, University of Michigan, Ann Arbor, MI, USA  }
\affil[2]{\normalsize Mathematics and Computer Science Division, Argonne National Laboratory, Lemont 60439 IL, USA \vspace{-1cm}}

\date{}
\begin{document}
\maketitle

\begin{abstract}
We present a resolvent-based framework for estimating turbulent velocity fluctuations in the wake of a spanwise-periodic NACA0012 airfoil at Mach 0.3, Reynolds number 23,000, and an angle of attack of $6^{\circ}$. Building on the methodology of Jung \textit{et al}. (\textit{J. Fluid Mech.} 2025, vol. 0, A1), we extend the approach to the more complex regime of a turbulent wake, which involves three primary challenges: (1) globally unstable modes in the linearized Navier–Stokes operator, (2) multi-scale turbulent structures, and (3) high-dimensional datasets. To address these challenges, we employ a data-driven approach that constructs causal resolvent-based estimation kernels from cross-spectral densities obtained via large-eddy simulations. These kernels are derived using the Wiener–Hopf method, which optimally enforces causality, thereby enhancing real-time estimation accuracy. The framework captures the spectral signatures of coherent structures and incorporates the colored statistics of nonlinear forcing terms acting on the linear system. To handle the computational demands of the high-dimensional estimation problem, we utilize parallel algorithms developed within the same framework. We further investigate sensor placement by analyzing single-sensor estimation error and coherence with target flow quantities. Results demonstrate accurate
causal estimation of streamwise velocity for the spanwise-averaged, spanwise-Fourier-transformed, and mid-span flow using limited shear-stress measurements on the surface of the airfoil. This study underscores the potential of the resolvent-based framework for efficient estimation in compressible, turbulent environments. \\  
\end{abstract}


\section{Introduction}\label{sec:intro}
\indent The estimation of turbulent wakes has been a longstanding focus in fluid mechanics, with decades of experimental and numerical studies revealing their complex dynamics across a range of conditions \citep{gartshore_two-dimensional_1967, reynolds_large-scale_1972, ghaemi_counter-hairpin_2011, shamsoddin_turbulent_2017, gupta_two-_2023, schauerte_experimental_2024}. At moderate Reynolds numbers (e.g., $Re \approx 20,000$) and mid-range angles of attack ($6^\circ$ -- $10^\circ$), a turbulent wake forms downstream of a laminar separation bubble and exhibits vortex shedding near a characteristic Strouhal number ($St_{\alpha} \approx 0.2$) based on airfoil thickness \citep{ducoin_numerical_2016, yeh_resolvent-analysis-based_2019}. Despite the complexity of these flows, even non-physics-based models have successfully estimated key features using experimental data alone \citep{hocevar_turbulent-wake_2005}. These results motivate the development of estimation methods that are not only accurate and interpretable but also suitable for real-time applications.\\
\indent Among model-based approaches, resolvent analysis has emerged as a powerful tool for studying and estimating turbulent flow structures. By interpreting the linearized Navier–Stokes equations as a forced system, resolvent analysis identifies the input–output relationships associated with the most amplified flow responses in the frequency domain \citep{jovanovic_componentwise_2005, mckeon_critical-layer_2010, towne_spectral_2018}. This framework has been used to guide both open-loop and closed-loop control designs. For instance, \citet{yeh_resolvent-analysis-based_2019} applied resolvent analysis to identify control parameters for separation control for a NACA airfoil at $Re=23,000$, 
$Ma=0.3$, and angles of attack of $6^{\circ}$ and $9^{\circ}$. Similarly, \citet{liu_unsteady_2021} reduced pressure fluctuations over a cavity using resolvent-informed open-loop control, while \citet{jin_optimal_2022} developed a resolvent-based iterative algorithm for closed-loop estimation and control in cylinder flow.\\
\indent More recently, resolvent-based estimation techniques have shown strong promise in capturing coherent structures in turbulent flows \citep{towne_resolvent-based_2020, martini_resolvent-based_2020, amaral_resolvent-based_2021, do_amaral_large-eddy-simulation-informed_2023, ying_resolvent-based_2023}. These methods leverage the close connection between resolvent modes and spectral proper orthogonal decomposition (SPOD) modes \citep{towne_spectral_2018}, enabling data-informed modeling. However, existing implementations have predominantly relied on non-causal formulations, limiting their utility for real-time estimation and control. To address this, recent work developed causal resolvent-based estimators using the Wiener–Hopf formalism \citep{noble1958}, which enforces causality \citep{daniele_fredholm_2007,martinelli_feedback_nodate} and yields optimal estimation kernels suited for closed-loop control \citep{martini_resolvent-based_2022} within the incompressible solver Nek5000 \citep{nek5000}. Extending this approach, \citet{jung_resovlent_based_2025_JFM} developed an efficient computational implementation integrated into the compressible solver CharLES \citep{bres_unstructured_2017}, successfully validating causal estimation and control methods for a laminar wake flow. Within the same resolvent-CharLES framework, \citet{towne_resolvent-based_2024} demonstrated the accuracy of linearized Navier–Stokes operators for various flows, leveraging methods proposed by \citet{nielsen_efficient_2006} and \citet{rutvij}.\\
\indent While resolvent analysis is naturally suited to globally stable flows, its extension to globally unstable systems presents challenges. In such cases, the relevance of the resolvent operator becomes less clear, since it is nominally associated with the steady-state response of the system \citep{duraisamy_chapter_2025}. One strategy to mitigate this involves exponential discounting to shift unstable eigenvalues into the stable region of the spectrum \citep{yeh_resolvent_2020}. Alternatively, data-driven resolvent formulations have been proposed to bypass the reliance on a globally stable linear operator \citep{martini_resolvent-based_2022, jung_resovlent_based_2025_JFM}. These methods maintain the benefits of resolvent-based estimation, specifically by remaining robust against global flow instability.\\
\indent Sensor placement is crucial for accurate flow estimation, influencing the fidelity of sensor-based measurements and subsequent control strategies. While formal optimization of sensor placement can significantly enhance estimation accuracy \citep{Manohar_data-driven_2018,Palash_DataDriven_2021,jin_optimal_2022}, such approaches often face practical limitations due to physical and operational constraints. Consequently, this study does not explicitly aim to optimize realistic sensor placement. Instead, effective sensor locations are determined empirically by assessing coherence between individual sensor–target pairs and by investigating single-sensor estimation errors, thereby providing practical insights relevant to realistic experimental scenarios. Within resolvent-based frameworks, spatial coherence has emerged as a particularly valuable metric for evaluating the suitability of sensor–target pairs. Recent experimental studies \citep{maia_real-time_2021,audiffred_experimental_2023,audiffred_reactive_2024} and numerical analyses \citep{towne_resolvent-based_2024,jung_resovlent_based_2025_JFM} demonstrate that coherence-based approaches effectively identify regions that yield accurate estimation and control. Furthermore, analyzing single-sensor estimation errors offers direct guidance for selecting sensor locations, particularly in high-coherence regions or along predictable disturbance propagation paths \citep{jung_resovlent_based_2025_JFM}. Alternatively, streamline visualizations can effectively identify sensor placements dynamically connected to target regions in convectively dominated flows \citep{Jung_RSV}.\\
\indent The objective of the present study is to estimate unsteady turbulent fluctuations in the wake of a NACA 0012 airfoil at Re = 23,000 using limited measurements on the surface of the airfoil.  Building on our prior work on a laminar wake \citep{jung_resovlent_based_2025_JFM}, we employ a causal resolvent-based estimation approach and focus on estimating velocity fluctuations in the turbulent wake, for which the flow exhibits broadband, multi-scale dynamics rather than periodic shedding. Unlike the laminar case, no upstream forcing is required to induce chaotic fluctuations; instead, we estimate naturally occurring turbulent fluctuations. The linearized Navier–Stokes operator in this regime is globally unstable, making traditional resolvent formulations poorly suited. To address this, we adopt a data-driven approach that constructs transfer functions from nonlinear simulation data, using empirical cross-spectral densities to approximate the resolvent operator \citep{martini_resolvent-based_2022, jung_resovlent_based_2025_JFM}. These estimators leverage the statistical structure of colored-in-time nonlinear forcing \citep{towne_resolvent-based_2020} and can be constructed without a priori model reduction, maintaining computational efficiency. Furthermore, we enforce causality through a Wiener–Hopf-based formalism \citep{noble1958}, enabling real-time estimation for control applications \citep{Jung2020,martini_resolvent-based_2022}. This work demonstrates that resolvent-based methods can provide accurate, interpretable, and scalable estimators for spanwise-periodic turbulent wakes by exploiting coherent flow structures \citep{towne_spectral_2018}.\\
\indent The remainder of this paper is organized as follows. In \S\ref{sec_Problem_setup}, we describe the flow configuration, numerical simulations, validation procedures, and the dataset used to construct data-driven estimation kernels, including definitions of three cases: spanwise-averaged, spanwise-Fourier, and mid-span plane. In \S\ref{sec_Methodology}, we briefly outline the resolvent-based estimation framework, including the theoretical formulation, the derivation of the causal resolvent-based estimation kernels via the Wiener-Hopf method, and implementation of the resolvent-based estimation tools. Estimation results for the three cases are discussed in detail in \S\ref{sec_Resolvent-basedEstimation}. Finally, concluding remarks and future research directions are presented in \S\ref{conclusion}.

\section{Problem set-up}\label{sec_Problem_setup}
\subsection{Problem description}\label{subsec_Problem_description}
We consider the turbulent wake of a NACA0012 airfoil at a moderate chord-based Reynolds number of $\mathrm{Re}_{L_{c}} = 23{,}000$ and an angle of attack of $\alpha = 6^\circ$, as studied in several previous works \citep{kojima_large-eddy_2013-1, munday_effects_2018, yeh_resolvent-analysis-based_2019, towne_database_2023}. The free-stream Mach number is set to $M_{\infty}\equiv u_{\infty}/c_{\infty}=0.3$, where $u_{\infty}$ and $c_{\infty}$ denote the free-stream velocity and sound speed, respectively. Shear-stress sensors are placed on the airfoil surface, while the targets are located downstream in the turbulent wake (see §\ref{sec_Resolvent-basedEstimation} for details). In this paper, we present a resolvent-based estimation of the time-series fluctuations in the streamwise velocity $u_{x}^{'}$ around the mean flow $\bar{u}_{x}$.
\subsection{Simulation}\label{subsec_Simulation}
\begin{figure}[t]
  \begin{center}
      \begin{tikzpicture}[baseline]
        \tikzstyle{every node}=[font=\small]
        \tikzset{>=latex}
        \node[anchor=south west,inner sep=0] (image) at (0,0) {
          \includegraphics[scale=1,width=1\textwidth]{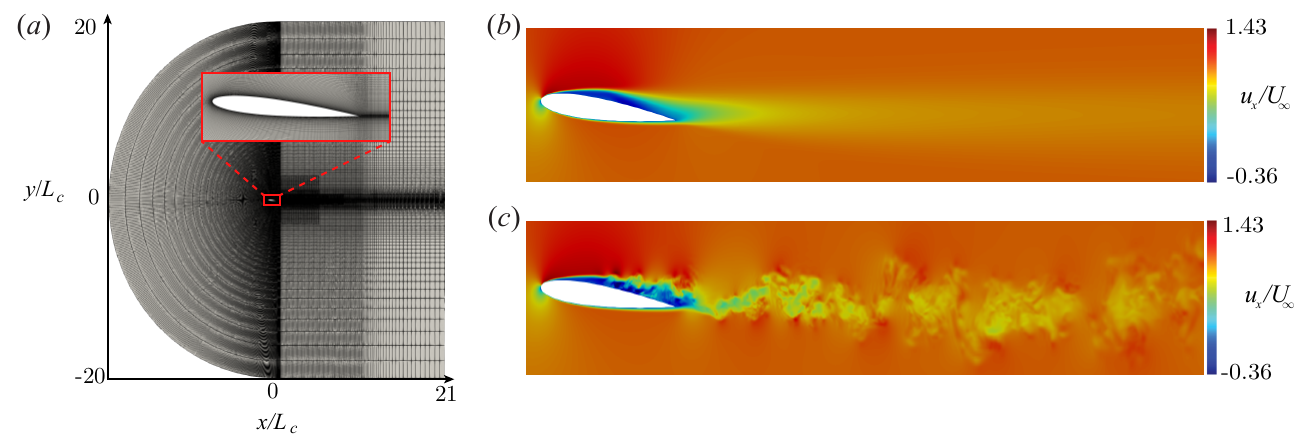}
        };
      \end{tikzpicture}
    \caption[]{\label{Fig_MESH} Computational mesh and simulation: (\textit{a}) the complete C-shaped grid, with a zoomed-in view of the mesh near the wall shown in the inset; (\textit{b}) the mean streamwise velocity $u_{x}$; and (\textit{c}) a snapshot of the instantaneous streamwise velocity $u_{x}$ obtained from the large-eddy simulation.}
  \end{center}
\end{figure}
\begin{figure}[t]
  \begin{center}
      \begin{tikzpicture}[baseline]
        \tikzstyle{every node}=[font=\small]
        \tikzset{>=latex}
        \node[anchor=south west,inner sep=0] (image) at (0,0) {
          \includegraphics[scale=0.8,width=0.8\textwidth]{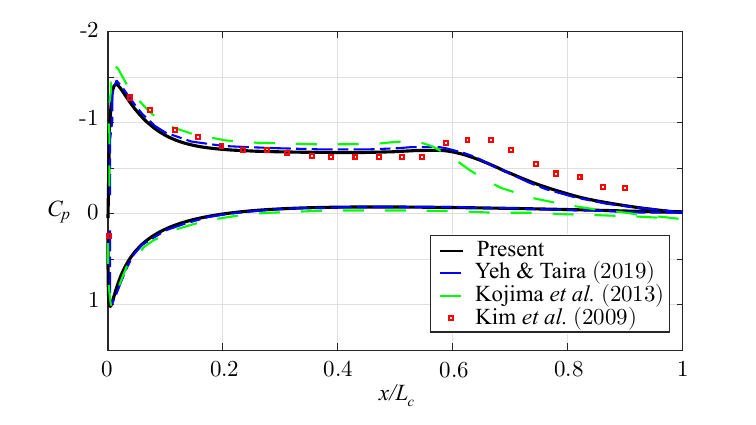}
        };
      \end{tikzpicture}
    \caption[]{\label{Fig_Cp} Comparison of the pressure coefficient $C_{p}$ with previous works.}
  \end{center}
\end{figure}

A large-eddy simulation (LES) utilizing the high-fidelity compressible flow solver CharLES from Cadence Design Systems \citep{bres_unstructured_2017,bres_importance_2018,yeh_resolvent-analysis-based_2019,bres2022gpu} is conducted to simulate turbulent airfoil flow. The simulation is conducted on Koehr, the Department of Defense's supercomputer, an HPE SGI 8600 system with a peak performance of 3.05 PFLOPS. For this study, 50 nodes of the supercomputer, each equipped with 48 cores, were utilized. A C-shaped mesh, generated using Pointwise, is depicted in figure \ref{Fig_MESH}(\textit{a}). Figures \ref{Fig_MESH}(\textit{b}) and (\textit{c}) illustrate the mean, instantaneous streamwise velocity field, respectively. The airfoil's leading edge is positioned at the origin, \(x/L_{c} = y/L_{c} = 0\). The computational domain spans \(x/L_{c} \in [-20, 21]\), \(y/L_{c} \in [-20, 20]\), and \(z/L_{c} \in [-0.1, 0.1]\). To optimize the mesh resolution, adaptive refinement was employed in the transverse and spanwise directions. A characteristic boundary condition \([\rho, u_{x}, u_{y}, u_{z}, P] = [\rho_{\infty}, U_{\infty}, 0, 0, P_{\infty}]\) was applied at the domain edges, with a sponge layer implemented for \(x/L_{c} \in [11, 21]\) to mitigate spurious reflections at the outflow. The time integration used a constant Courant-Friedrichs-Lewy (CFL) number of 0.84.\\
\indent We validate the LES by comparing the pressure coefficient and aerodynamic forces 
\begin{equation}\label{eq1}
C_{p}=\frac{p-p_{\infty}}{\frac{1}{2} \rho_{\infty} U_{\infty}^{2} A},\quad C_{D}=\frac{F_{D}}{\frac{1}{2} \rho_{\infty} U_{\infty}^{2} A},\quad \textrm{and} \quad C_{L}=\frac{F_{L}}{\frac{1}{2} \rho_{\infty} U_{\infty}^{2} A}
\end{equation}
with previous works in table \ref{tab_CdCl} and figure \ref{Fig_Cp}. Experimental data is provided by \citet{kim2009}. The closest analogue to our simulation is that of \citet{yeh_resolvent-analysis-based_2019}, which entails the same physical setup and solver but a different grid. The simulations of \citet{kojima_large-eddy_2013-1} and \citet{munday_effects_2018} differ due to their use of an incompressible flow solver.
\begin{table}[t]
\centering
 \begin{tabular}{ c c c }
\hline
& $\bar{C}_{D}$ & $\bar{C}_{L}$\\
\hline
Present study & 0.0663 & 0.5836\\
\citet{yeh_resolvent-analysis-based_2019} & 0.066 & 0.609\\
\citet{munday_effects_2018} & 0.062 & 0.637\\
 \citet{kojima_large-eddy_2013-1} & 0.054 & 0.639 \\

 \hline

\end{tabular}
\caption{Comparison of the time-averaged drag and lift coefficients with the reference data for a NACA 0012 airfoil at $\emph{Re}_{L_{c}} = 23,000$ and angle of attack $6^{\circ}$.}
\label{tab_CdCl}
\end{table}

\begin{figure}[t]
  \begin{center}
      \begin{tikzpicture}[baseline]
        \tikzstyle{every node}=[font=\small]
        \tikzset{>=latex}
        \node[anchor=south west,inner sep=0] (image) at (0,0) {
          \includegraphics[scale=1,width=1\textwidth]{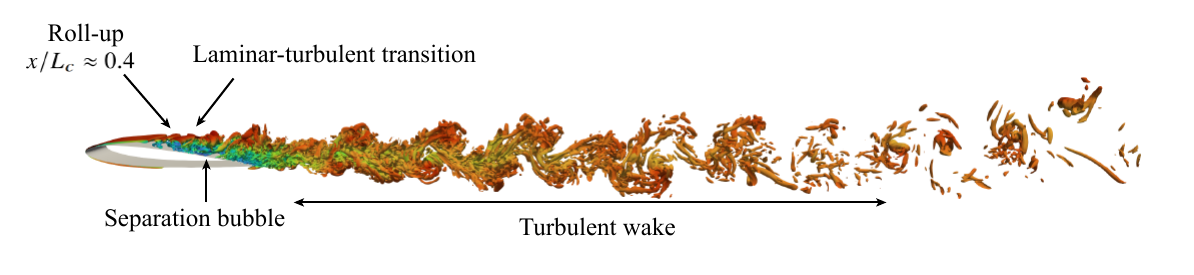}
        };
      \end{tikzpicture}
    \caption[]{\label{Fig_Qcriterion} Q-criterion visualization showing the key flow features.}
  \end{center}
\end{figure}

As shown in figure \ref{Fig_Qcriterion}, the shear layer roll-up on the suction surface occurs at $x/L_{c}\approx 0.4$, leading to spanwise vortices. A laminar separation bubble is observed on the suction side of the airfoil in the region $0.1 \lesssim x/L_{c} \lesssim 0.84$, as indicated by the mean streamwise velocity shown in figure \ref{Fig_MESH}(\textit{b}). Laminar-turbulent transition is detected at $x/L_{c}\approx 0.6$, as determined by the termination of the pressure coefficient plateau illustrated in figure \ref{Fig_Cp}. Finally, a turbulent wake develops behind the airfoil.

\subsection{Dataset}
\indent We consider three distinct representations of the spanwise-periodic flow: spanwise-averaged, spanwise-Fourier, and mid-span plane cases. The spanwise-averaged case is the simplest representation, defined mathematically as \begin{equation}\label{eq4.0} {\boldsymbol{q}}_{\text{span-avg}}(x,y,t) = \frac{1}{L_z} \int_{0}^{L_z} \boldsymbol{q}(x,y,z,t) \, dz, \end{equation} where $L_z$ is the spanwise length. Spanwise-Fourier decomposition, a common technique in biglobal stability analyses of three-dimensional flows \citep{theofilis_advances_2003}, has been previously utilized to investigate resolvent gains at various spanwise wavenumbers \citep{yeh_resolvent-analysis-based_2019, towne_database_2023}. We similarly analyze spanwise-Fourier modes defined by
\begin{equation}\label{eq4.1} \hat{\boldsymbol{q}}(x,y,k_{z},t) = \frac{1}{L_z}\int_{0}^{L_z} \boldsymbol{q}(x,y,z,t)e^{-ik_{z}z} \, dz, \end{equation} where $k_{z}$ denotes the spanwise wavenumber. This approach allows us to progressively incorporate more complex flow structures by examining the spanwise Fourier modes. Finally, the mid-span plane case focuses on the flow at the central spanwise location, defined as \begin{equation}\label{eq4.2} \boldsymbol{q}_{\text{mid}}(x,y,t) = \boldsymbol{q}(x,y,z_{\text{mid}},t), \end{equation} where $z_{\text{mid}} = L_z/2$ is the mid-span location. This representation helps capture spanwise effects in the estimation process by examining the flow at a representative spanwise position and is consistent with a realistic use case where sensors are distributed along the airfoil at a single spanwise position. These representations enhance our understanding of spanwise effects in the flow and support investigations into sensor placement and flow estimation strategies for a spanwise-periodic flow.\\
\indent The dataset comprises \(n_t = 75{,}000\) time-resolved, three-dimensional snapshots of \([\rho, \rho u_x, \rho u_y, \rho u_z, T]\), spanning a nondimensional time window of \(tU_\infty / L_c \in [0, 324]\). These snapshots are sampled at a uniform interval of \(\Delta t U_\infty / L_c = 0.00432\), which resolves the dominant vortex-shedding frequency at \(St_{\alpha} \equiv f L_c (sin \alpha) / U_\infty \approx 0.15\). This finer temporal resolution and longer duration compared to previous datasets \citep{towne_database_2023} enable capturing smaller-scale turbulent fluctuations in time and provide a larger number of snapshots, resulting in improved convergence of the estimation kernels. The time window captures approximately 466 shedding cycles, providing a sufficiently long time series to ensure statistical convergence of spectral quantities, including the cross-spectral densities used in constructing the estimation kernel. The spatial domain covers \(x / L_c \in [-0.5, 2.5]\) and \(y / L_c \in [-0.5, 0.5]\), capturing the near-wake region with high spatial fidelity.\\ 
\indent To construct resolvent-based estimation kernels, we follow the implementation of \citet{jung_resovlent_based_2025_JFM}. Prior to estimation, the dataset was preprocessed by subtracting the temporal mean from each variable and applying a Hamming window \citep{hamming1997digital} to reduce spectral leakage in the Fourier transform. For assessing generalization performance, 80\% of the data is used to train the estimation model, while the remaining 20\% is reserved for testing.
\section{Resolvent-based estimation framework}\label{sec_Methodology}
In this section, we briefly review the resolvent-based estimation framework developed in recent studies \citep{towne_resolvent-based_2020, martini_resolvent-based_2020, martini_resolvent-based_2022, jung_resovlent_based_2025_JFM}.\\
\subsection{Estimation system}\label{Estimation_system}
We start with the compressible Navier-Stokes equations written as
\begin{equation}\label{eq2}
\frac{\partial \boldsymbol{q}}{\partial t}= \mathcal{F} (\boldsymbol{q}),
\end{equation}
where $\boldsymbol{q}$ is a state vector of flow variables $[\rho,\rho u_{x},\rho u_{y},\rho u_{z},\rho E]^{T}$ and $\mathcal{F}$ is the nonlinear Navier-Stokes operator. The equations are linearized using a Reynolds decomposition, giving

\begin{equation}\label{eq3}
\frac{\partial \boldsymbol{q}'}{\partial t}-A\boldsymbol{q'}=\boldsymbol{f}(\boldsymbol{\bar{q}},\boldsymbol{q'}),
\end{equation}
where $\bar{\boldsymbol{q}}$ and $\boldsymbol{q}'$ represent the mean and perturbation state vectors of the flow variables, respectively, $\boldsymbol{A} = \frac{\partial \mathcal{F}}{\partial \boldsymbol{q}}\big|_{\bar{q}}$ is the linearized Navier-Stokes operator, and  $\boldsymbol{f}$ represents the remaining nonlinear terms. For convenience, we drop the $(\cdot )^{'}$ superscript for perturbations from this point on.\\
\indent To derive the estimation method, we consider a generalization of (\ref{eq2}) in the form of the linear-time-invariant (LTI) system

\begin{subequations}
\begin{alignat}{2}\label{eq_LinearSystem}
\frac{d\boldsymbol{q}}{dt}(t)&=
\mathsfbi{A}\boldsymbol{q}(t)+\mathsfbi{B}_{f}\boldsymbol{f}(t),\\\label{eq_LinearSystem_y}
\boldsymbol{y}(t)&=
\mathsfbi{C}_{y}\boldsymbol{q}(t)+\boldsymbol{n}(t),\\\label{eq_LinearSystem_z}
\boldsymbol{z}(t)&=
\mathsfbi{C}_{z}\boldsymbol{q}(t),
\end{alignat}
\end{subequations}
where the forcing matrix $\mathsfbi{B}_{f} \in \rm \mathbb{C}^{n \times n_{f}}$ can be used to restrict the form of the forcing $\boldsymbol{f}$, and the measurement $\boldsymbol{y}$ and target $\boldsymbol{z}$ indicate readings of the state perturbation extracted by a measurement matrix $\mathsfbi{C}_{y}\in \rm \mathbb{C}^{n_{y} \times n}$ and a target matrix $\mathsfbi{C}_{z}\in \rm \mathbb{C}^{n_{z} \times n}$. The number of sensors and targets is denoted ${n}_{y}$ and $n_{z}$, respectively. The vector $\boldsymbol{n}$ indicates the sensor noise. The measurement $\boldsymbol{y}$ is chosen to be the shear stress $\boldsymbol{\tau}_{w}$ at one or more points on the surface of the airfoil. It is computed from the state $\boldsymbol{q}$ using a suitably defined measurement matrix $\mathsfbi{C}_{y}$, as in \citet{jung_resovlent_based_2025_JFM}. In line with recent applications of the resolvent-based framework \citep{amaral_resolvent-based_2021, towne_resolvent-based_2024}, preliminary tests indicated minimal difference between using shear stress and pressure as the measured quantity on the airfoil surface. The target $\boldsymbol{z}$ represents one or more Gaussian sensors, each with spatial support given by
\begin{equation}\label{eq7}
\alpha e^{-(x-x_{c})^{2}/ 2 \sigma_{x} ^{2} -(y-y_{c})^{2}/ 2 \sigma_{y} ^{2}-(z-z_{c})^{2}/ 2 \sigma_{z} ^{2}},
\end{equation}
where $\sigma_{x}$, $\sigma_{y}$, and $\sigma_{z}$ define the width of the Gaussian distribution in each coordinate direction, and $\alpha$ is a weighting factor representing the integrated contribution over the Gaussian support. These targets are extracted from the state using the measurement matrix $\mathsfbi{C}_{z}$. \\
\indent By Fourier-transforming the linear system into the frequency domain and defining the resolvent operator as $\mathsfbi{R} = (i \omega \mathsfbi{I} - \mathsfbi{A})^{-1}$, where $\mathsfbi{I}$ is the identity matrix and $\omega$ is the frequency, we derive modified resolvent operators used for constructing the resolvent-based kernels for estimation. The measurement and target vectors can be written as 

\begin{subequations}
\begin{alignat}{2}\label{eq_yhat}
\boldsymbol{\hat{y}}&=\mathsfbi{R}_{yf}\boldsymbol{\hat{f}}+\boldsymbol{\hat{n}},\\\label{eq_zhat}
\boldsymbol{\hat{z}}&= \mathsfbi{R}_{zf}\boldsymbol{\hat{f}}
\end{alignat}
\end{subequations}
with $\mathsfbi{R}_{yf}=\mathsfbi{C}_{y}\mathsfbi{R}\mathsfbi{B}_{f}$, and
$\mathsfbi{R}_{zf}=\mathsfbi{C}_{z}\mathsfbi{R}\mathsfbi{B}_{f}$. The notation ($\hat{\cdot }$) indicates a quantity in the frequency domain.

\subsection{Kernels for resolvent-based estimation}\label{subsec_Resolvent-based estimation}
\begin{figure}[t]
  \begin{center}
      \begin{tikzpicture}[baseline]
        \tikzstyle{every node}=[font=\small]
        \tikzset{>=latex}
        \node[anchor=south west,inner sep=0] (image) at (0,0) {
          \includegraphics[scale=1,width=1\textwidth]{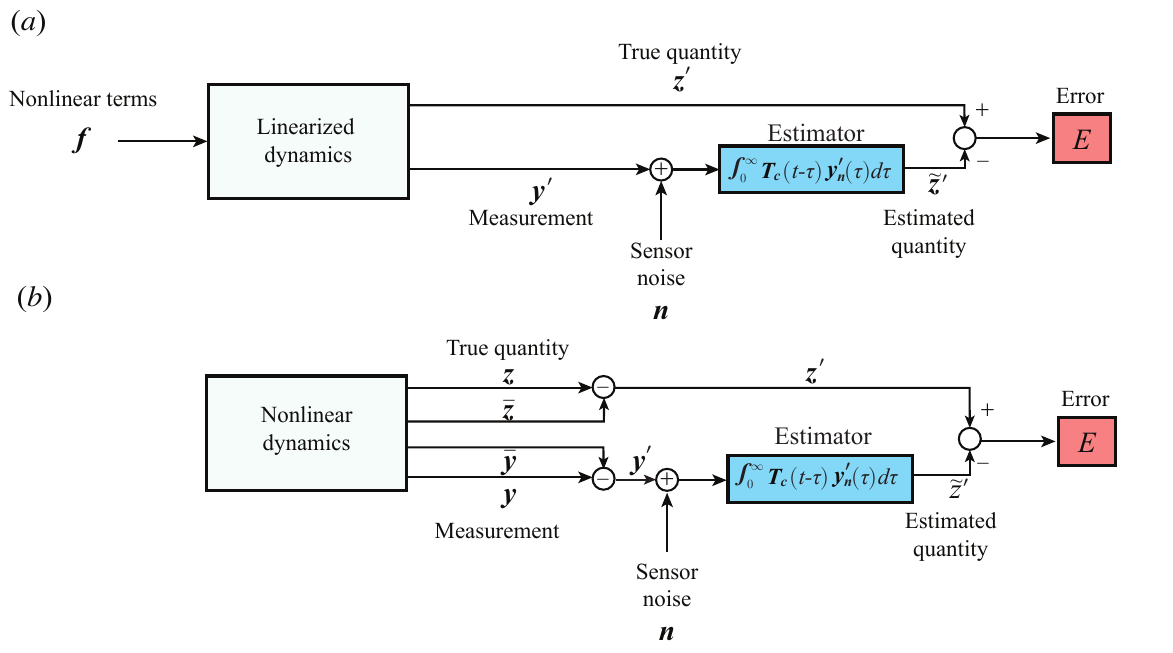}
        };
      \end{tikzpicture}
    \caption[]{\label{Fig_Block} Block diagram of resolvent-based estimation: (\textit{a}) design of the resolvent-based estimator based on the linear system and (\textit{b}) application to the nonlinear system. Here, $(\cdot)'$ indicates fluctuations around the mean quantity, $(\bar{\cdot})$ denotes mean quantity, and $(\tilde{\cdot})$ represents an estimated quantity.}
  \end{center}
\end{figure}
\indent We perform estimation with three different kernels: non-causal, truncated non-causal, and causal \citep{martini_resolvent-based_2022}. First, we start with a non-causal estimator using a convolution function, written as
\begin{equation}\label{eq10}
\Tilde{\boldsymbol{z}}_{nc}(t)=\int_{-\infty}^{\infty} \mathsfbi{T}_{nc}(t-\tau) \ \boldsymbol{y}_{n}(\tau)d\tau,
\end{equation}
where $\mathsfbi{T}_{nc} \in \rm I\!R^{n_{z} \times n_{y}}$ is a non-causal estimation kernel between the sensor measurement and the estimated target state. The objective is to select the kernel to minimize the expected value of the estimation error $e(t)=\Tilde{\boldsymbol{z}}-\boldsymbol{z}(t)$, and this is formalized by defining the cost function
\begin{equation}\label{eq11}
\mathsfbi{J}_{nc}(t)=\int_{-\infty}^{\infty} \mathbb{E} \{ e(t)^{\dagger}  e(t) \} \ dt,
\end{equation}
where $\mathbb{E} \{ \cdot \}$ is the expectation operator. The overall goal can be explained more easily using the block diagram in figure \ref{Fig_Block}(\textit{a}).

By minimizing the cost function, we obtain a non-causal estimation kernel
\begin{equation}\label{eq_Tnc}
\mathsfbi{\hat{T}}_{nc}(\omega)=  (\mathsfbi{R}_{zf} \mathsfbi{\hat{F}} \mathsfbi{R}_{yf}^{\dagger} )(\mathsfbi{R}_{yf} \mathsfbi{\hat{F}} \mathsfbi{R}_{yf}^{\dagger}  + \mathsfbi{\hat{N}})^{-1} ,
\end{equation}
where $\mathsfbi{\hat{F}} = \mathbb{E} \{ \hat{\boldsymbol{f}} \hat{\boldsymbol{f}}^{\dagger} \}$ and $\mathsfbi{\hat{N}} = \mathbb{E} \{ \hat{\boldsymbol{n}}  \hat{\boldsymbol{n}}^{\dagger} \}$ denote the forcing and sensor noise cross-spectral density matrices, respectively. Note that the space-time color of the forcing terms is explicitly included in (\ref{eq_Tnc}), which means that the nonlinearity of the flow can be statistically accounted for in the transfer function.

\indent Second, for real-time estimation, the only current and past (but not future) sensor measurements are available to the estimator. One common approach to addressing this practical limitation is to truncate the non-causal kernel by setting the part associated with unavailable future measurements to zero. The truncated non-causal kernel is defined as
\begin{eqnarray}\label{eq16}
    \mathsfbi{T}_{tnc}(\tau)= 
\begin{cases}
    \mathsfbi{T}_{nc}(\tau),& \tau\geq 0,\\
    0,              & \tau<0.
\end{cases}
\end{eqnarray}
The truncated non-causal estimates are computed as
\begin{equation}\label{eq17}
\Tilde{\boldsymbol{z}}_{tnc}(t)=\int_{-\infty}^{0} \mathsfbi{T}_{tnc}(t-\tau) \ \boldsymbol{y}_{n}(\tau)d\tau.
\end{equation}
\indent Lastly, by enforcing causality via Wiener-Hopf formalism in cost functions, we derive optimal causal estimation kernels under the constraint of causality. A causal estimator 
\begin{equation}\label{eq13}
\Tilde{\boldsymbol{z}}_{c}(t)=\int_{-\infty}^{0} \mathsfbi{T}_{c}(t-\tau) \ \boldsymbol{y}_{n}(\tau)d\tau,
\end{equation}
is defined in terms of the causal estimation kernel $\mathsfbi{T}_{c} \in \rm I\!R^{n_{z} \times n_{y}}$. To enforce causality, we modify the cost function of (\ref{eq11}) to read
\begin{equation}\label{eq14}
\mathsfbi{J}_{c}(t)=\int_{-\infty}^{\infty} \mathbb{E}\{ e(t)^{*}  e(t)\} + (\mathsfbi{\Lambda}_{-}(t) \mathsfbi{T}_{c} (t) + \mathsfbi{\Lambda}_{-}^{\dagger}(t) \mathsfbi{T}_{c}^{\dagger} (t)) \ dt,
\end{equation}
where $\mathsfbi{\Lambda}$ is a Lagrange multiplier that is used to force the causal kernel to be zero for the non-causal part ($\tau<0$). The $(\cdot)_{+}$ and $(\cdot)_{-}$ subscripts represent the non-causal ($\tau<0$) and causal ($\tau>0$) parts of any matrix or function to be zero by achieving a Wiener-Hopf factorization. Additionally, $[\cdot]_{+}$ and $(\cdot)_{+}$ denote additive and multiplicative factorization operators, respectively (see Appendix~\ref{app_WH} for further details). The derived causal kernel for the estimation is
\begin{eqnarray}\label{eq_Tc}
\mathsfbi{\hat{T}}_{c}(\omega) &=&  [ \mathsfbi{{R}}_{zf} \mathsfbi{\hat{F}} \mathsfbi{{R}}_{yf}^{\dagger} (\mathsfbi{{R}}_{yf} \mathsfbi{\hat{F}} \mathsfbi{R}_{yf}^{\dagger} + \mathsfbi{\hat{N}})^{-1}_{-}]_{+}(\mathsfbi{{R}}_{yf} \mathsfbi{\hat{F}} \mathsfbi{R}_{yf}^{\dagger} + \mathsfbi{\hat{N}} )^{-1}_{+}.
\end{eqnarray}

\indent All the transfer functions we consider involve the forcing cross-spectra density matrix that can account for the impact of the nonlinearity in a statistical sense. However, our method is more suitable for the real-time estimation approach since causality is constrained. In this paper, we present the results for both the causal and truncated non-causal transfer functions to assess how optimally enforcing causality impacts the estimation accuracy for turbulent flow.

The resolvent operators appearing in the estimation kernels are not well-defined for globally unstable systems \citep{jovanovic_componentwise_2005,yeh_resolvent_2020} such as the linearization of the turbulent airfoil wake about its mean \citep{yeh_resolvent-analysis-based_2019}. To circumvent this issue, we use a data-driven approach \citep{martini_resolvent-based_2022} that enables the construction of resolvent-based kernels without the need to explicitly form resolvent operators. The empirical cross-spectra densities (CSD) from the simulation are used to obtain the modified resolvent operators, which is a similar way to the previous work \citep{audiffred_experimental_2023,jung_resovlent_based_2025_JFM}. Following the notation of the LTI system in $\S$\ref{Estimation_system}, the datasets of the sensor and target readings are Fourier-transformed, expressed as

\begin{equation}\label{eq18}
\begin{bmatrix}
\hat{\boldsymbol{y}}\\
\hat{\boldsymbol{z}}
\end{bmatrix}
=
\begin{bmatrix}
\mathsfbi{R}_{y,f} & 1  \\
\mathsfbi{R}_{z,f} & 0 
\end{bmatrix}
\begin{bmatrix}
\hat{\boldsymbol{f}} \\
\hat{\boldsymbol{n}} 
\end{bmatrix}
.
\end{equation}
 Computing the cross-spectral density of $\left[ \hat{\boldsymbol{y}} \quad \hat{\boldsymbol{z}} \right]^{T}$ gives  
\begin{equation}\label{eq_Syy1_Szy1}
\begin{bmatrix}
\mathsfbi{S}_{yf,yf} &\mathsfbi{S}_{yf,zf} \\
\mathsfbi{S}_{zf,yf}  & \mathsfbi{S}_{zf,zf} 
\end{bmatrix}
=
\begin{bmatrix}
\mathsfbi{R}_{yf} \mathsfbi{\hat{F}} \mathsfbi{R}_{yf}^{\dagger} &\mathsfbi{R}_{yf} \mathsfbi{\hat{F}} \mathsfbi{R}_{zf}^{\dagger}  \\
\mathsfbi{R}_{zf} \mathsfbi{\hat{F}} \mathsfbi{R}_{yf}^{\dagger}   & \mathsfbi{R}_{zf} \mathsfbi{\hat{F}} \mathsfbi{R}_{zf}^{\dagger}  
\end{bmatrix},
\end{equation}
with $\mathsfbi{S}_{yf,yf}=\mathbb{E} \{ \hat{\boldsymbol{y}} \hat{\boldsymbol{y}}^{*} \}$ and $\mathsfbi{S}_{zf,yf}=\mathbb{E} \{ \hat{\boldsymbol{z}} \hat{\boldsymbol{y}}^{*} \}$. Since the right-hand side of (\ref{eq_Syy1_Szy1}) contains the terms needed to build the estimation kernels, this shows that the correlations on the left-hand side can be used in their place in (\ref{eq_Tnc}) and (\ref{eq_Tc}). The data-driven non-causal and causal estimation kernels in (\ref{eq_Tnc}) and (\ref{eq_Tc}) are computed using the CSDs from (\ref{eq_Syy1_Szy1}), yielding
\begin{subequations}\label{eq_T_nc_c_D}
\begin{alignat}{2}\label{eq_T_nc_D}
\mathsfbi{\hat{T}}_{nc} &= \mathsfbi{S}_{zy} ( \mathsfbi{S}_{yy} + \mathsfbi{\hat{N}})^{-1},\\\label{eq_c_D}
\mathsfbi{\hat{T}}_{c} &= [\mathsfbi{S}_{zy} ( \mathsfbi{S}_{yy} + \mathsfbi{\hat{N}})^{-1}_{-}]_{+} (\mathsfbi{S}_{yy} + \mathsfbi{\hat{N}})_{+}^{-1}.
\end{alignat}
\end{subequations}

Note that the CSDs inherently contain statistical information related to the nonlinearities of the flow within the forcing covariance matrix. The mean flow is subtracted from both the measurement readings and target values. Estimation kernels are computed offline for the linear system, as illustrated in figure \ref{Fig_Block}(\textit{a}). These kernels are then used for real-time (online) estimation of velocity fluctuations in the wake, applying the approach to the nonlinear system, as depicted in figure \ref{Fig_Block}(\textit{b}).
\subsection{Implementation of resolvent-based estimation}
\begin{figure}[t]
  \begin{center}
      \begin{tikzpicture}[baseline]
        \tikzstyle{every node}=[font=\small]
        \tikzset{>=latex}
        \node[anchor=south west,inner sep=0] (image) at (0,0) {
          \includegraphics[scale=1,width=1\textwidth]{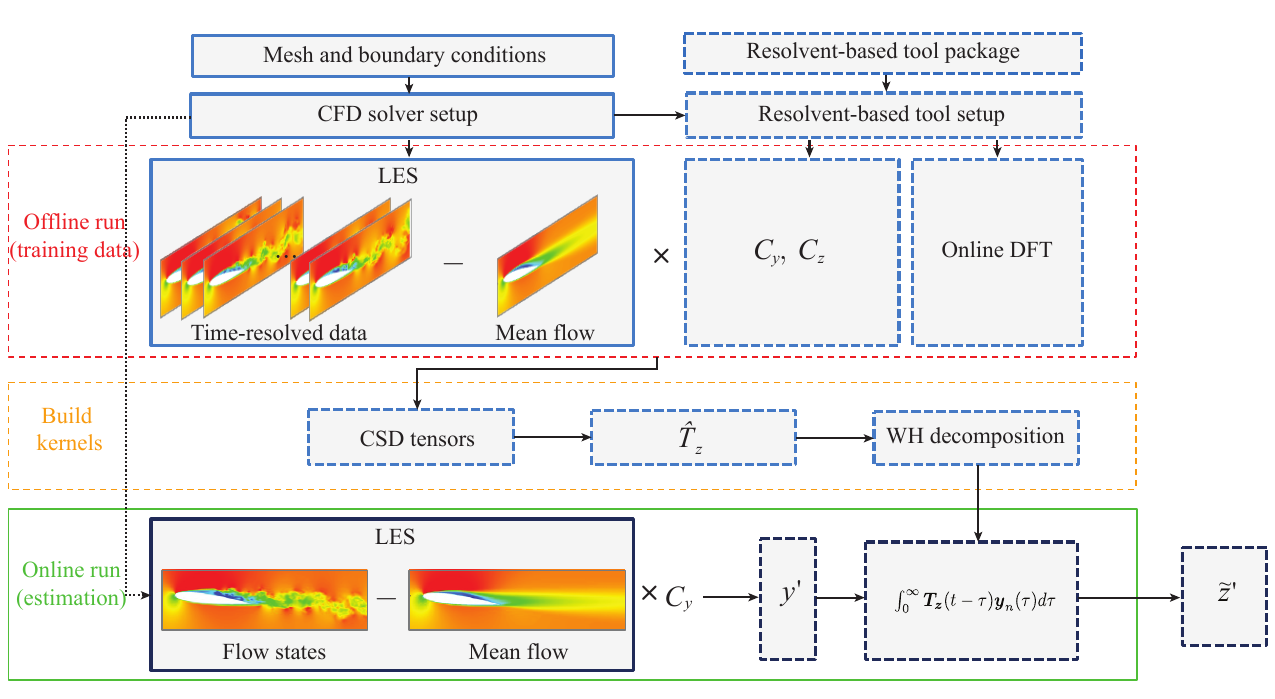}
        };
      \end{tikzpicture}
    \caption[]{\label{Fig_RSVtools} Flowchart of the resolvent-based estimation tool within a compressible flow solver. Dashed-line boxes represent the resolvent-based tool integrated into the existing CFD solver (solid-line box). Blue boxes, including those outlined in red (offline run) and orange (build kernels) dashed lines, indicate offline processes, while black boxes with green solid outlines represent online (estimation) processes.}
  \end{center}
\end{figure}
We employ recently developed resolvent-based estimation and control tools integrated within the compressible flow solver CharLES \citep{jung_resolvent-based_2024, towne_resolvent-based_2024}, which was carefully designed to scale to large-scale problems such as a turbulent wake. Originally inspired by the incompressible solver implementation by \citet{martini_resolvent-based_2022}, the approach was extended by \citet{jung_resovlent_based_2025_JFM} to compressible flows using parallel algorithms written in C++, built directly upon the existing solver architecture. These resolvent-based tools interact with the solver, enabling efficient in-situ computations during simulations. Moreover, our modular software design simplifies integration with other relevant software packages and ensures straightforward deployment in high-performance computing environments.\\
\indent An overview of the implementation is shown in figure~\ref{Fig_RSVtools}. Our compressible resolvent-based tool integrates external libraries such as PETSc \citep{PETSC2019} and FFTW \citep{frigo_design_2005}, enabling scalable parallel computation through in-situ matrix operations with flow quantities during the training process. The tool constructs measurement matrices (\(\mathsfbi{C}_{y}\) and \(\mathsfbi{C}_{z}\)) using control volume indices from the CFD mesh. For cases involving high-rank targets (\(n_{z} \gg n_{y}\)), an online discrete Fourier transform (DFT) is used to minimize memory consumption while preserving time-resolved information \citep{schmidt_efficient_2019, farghadan_efficient_2024, farghadan_scalable_2025}. Training data is obtained from large eddy simulations (LES), and the corresponding cross-spectral densities (CSDs) are used to generate the estimation kernels. The DFT parameters used to obtain converged kernels are described in Appendix \ref{app_Conv_kernels}.
\section{Results}\label{sec_Resolvent-basedEstimation}
\begin{figure}[t]
  \begin{center}
      \begin{tikzpicture}[baseline]
        \tikzstyle{every node}=[font=\small]
        \tikzset{>=latex}
        \node[anchor=south west,inner sep=0] (image) at (0,0) {
          \includegraphics[scale=1,width=1\textwidth]{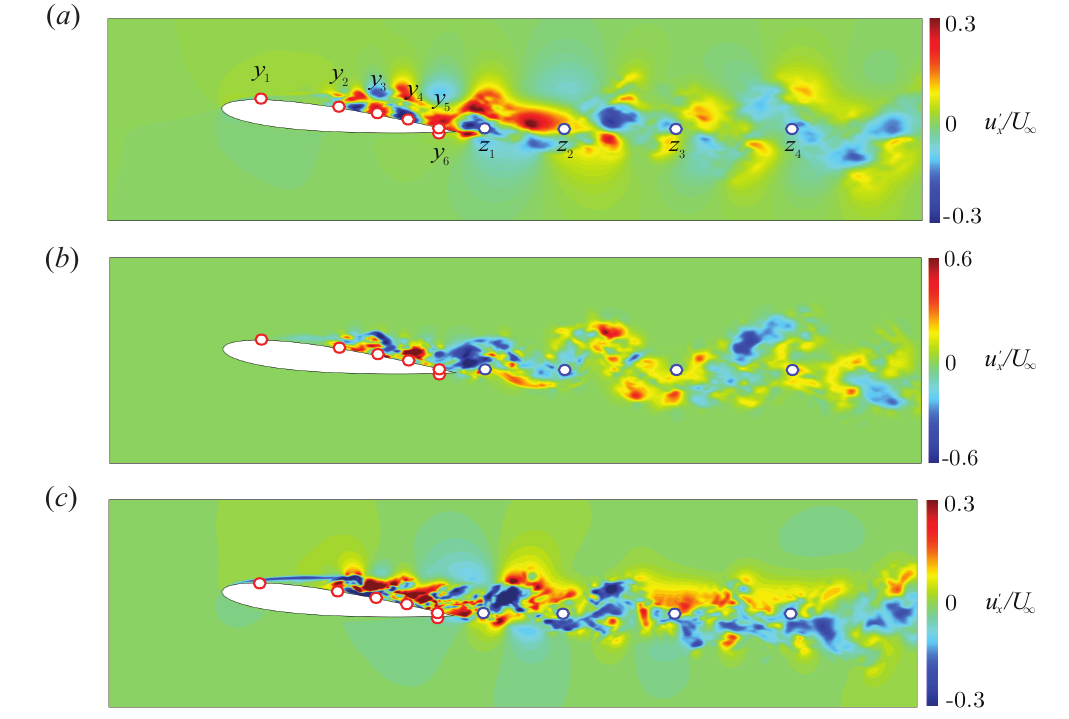}
        };
      \end{tikzpicture}
    \caption[]{\label{Fig_velfield_all} Streamwise velocity fluctuation fields ($u_{x}'/U_{\infty}$) for (\textit{a}) spanwise-average, (\textit{b}) spanwise-Fourier mode $k_{z}=10\pi$, and (\textit{c}) mid-span plane case. Target locations ($z_{1}, z_{2}, z_{3}, z_{4}$) are indicated by blue circles and positioned along the trailing edge at streamwise coordinates $x/L_{c} = [1.2,,1.5,,2.0,,2.5]$ for a constant transverse position $y/L_c = -0.11$.}
  \end{center}
\end{figure}
We demonstrate the effectiveness of the causal resolvent-based approach for predicting turbulent fluctuations over a NACA0012 airfoil at a $6^{\circ}$ angle of attack. Unlike earlier resolvent-based studies that employed non-causal estimation methods \citep{towne_resolvent-based_2020, martini_resolvent-based_2020, amaral_resolvent-based_2021, do_amaral_large-eddy-simulation-informed_2023}, the causal estimator uses only present and past measurements and kernels optimized for this purpose via the Wiener-Hopf method \citep{martini_resolvent-based_2022}. This causality enforcement makes our method particularly suitable for real-time applications. Significantly, this work is the first numerical investigation explicitly applying the Wiener-Hopf method to enforce causality in turbulent flow estimation \citep{jung_toward_2024}, extending previous efforts primarily focused on experimental flow-control studies \citep{audiffred_experimental_2023, audiffred_reactive_2024} and laminar-flow scenarios \citep{martini_resolvent-based_2022, jung_resovlent_based_2025_JFM}.\\
\indent As previously mentioned, our analysis centers on the spanwise-averaged, spanwise-Fourier, and mid-span plane cases, as depicted in figure~\ref{Fig_velfield_all}. The specific target locations analyzed in the wake region are also highlighted in figure~\ref{Fig_velfield_all}(\textit{a}). The spanwise-averaged case provides a concise yet informative view of spatial structure in the spanwise direction, enabling a focused analysis of sensor positioning in the streamwise and cross-stream planes. We progressively incorporate more complex flow structures by examining the spanwise-Fourier modes, with particular attention to the most energetic mode at $k_{z}=10\pi$, and the flow on the mid-span plane to better capture spanwise effects in the estimation process.\\
\indent Given the large number of potential sensor positions available on the three-dimensional airfoil surface, our investigation begins with selecting the most effective sensor locations based on single-sensor estimation error metrics and coherence analyses. Subsequently, we examine the power spectral density (PSD) at these selected sensor and target locations to better understand their frequency-domain characteristics. We then construct estimation kernels specifically designed for multi-sensor estimation, analyzing their characteristics in both the time and frequency domains to gain insight into their role in estimation performance. Finally, we present multi-sensor estimation results, quantifying errors at both single target points and over extended target regions, to demonstrate the accuracy of the resolvent-based method.\\

\subsection{Investigation of sensor placement}\label{sub_sensorplacement}
\begin{figure}[t]
  \begin{center}
      \begin{tikzpicture}[baseline]
        \tikzstyle{every node}=[font=\small]
        \tikzset{>=latex}
        \node[anchor=south west,inner sep=0] (image) at (0,0) {
          \includegraphics[scale=0.7,width=0.7\textwidth]{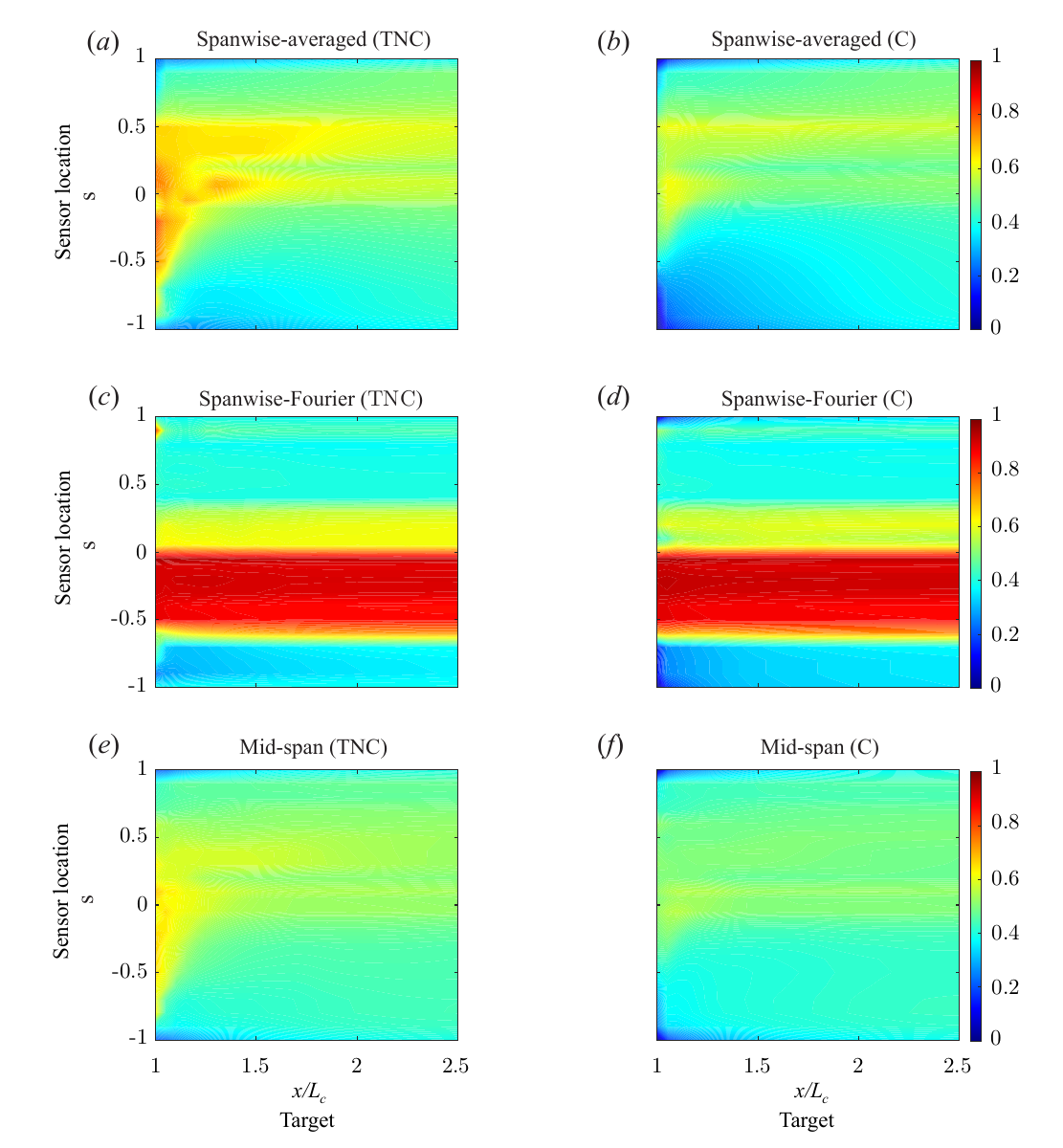}
        };
      \end{tikzpicture}
    \caption[]{\label{Fig_EST_Error_field} Estimation error maps for the streamwise velocity fluctuation $u_{x}'$ in single sensor-target configurations: (\textit{a}, \textit{b}) spanwise-averaged flow, (\textit{c}, \textit{d}) spanwise-Fourier mode for $k_z=10\pi$, and (\textit{e}, \textit{f}) mid-span plane flow. Results using the truncated non-causal approach are shown in panels (\textit{a}, \textit{c}, \textit{e}), while the causal estimation results are shown in panels (\textit{b}, \textit{d}, \textit{f}).}
  \end{center}
\end{figure}
\begin{figure}[t]
  \begin{center}
      \begin{tikzpicture}[baseline]
        \tikzstyle{every node}=[font=\small]
        \tikzset{>=latex}
        \node[anchor=south west,inner sep=0] (image) at (0,0) {
          \includegraphics[scale=1,width=1\textwidth]{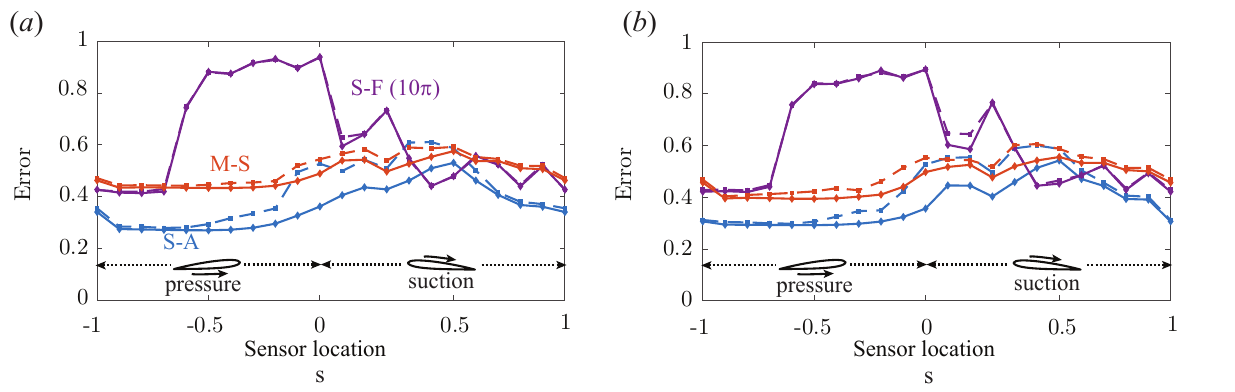}
        };
      \end{tikzpicture}
    \caption[]{\label{Fig_EST_generaltrend} Estimation errors of streamwise (\textit{a}) and cross-streamwise (\textit{b}) velocities as a function of the sensor location on the airfoil surface. The blue, orange, purple, and green lines represent spanwise-averaged (S-A), mid-span (M-S), and spanwise-Fourier (S-F) ($K_{z}=10\pi$) flows, respectively. The solid diamond ($\diamond$) and dashed square ($\square$) lines denote the causal and truncated non-causal approaches, respectively.}
  \end{center}
\end{figure}

\indent Sensor placement is guided by evaluating single-sensor estimation error metrics, similar to the approach in \citet{jung_resovlent_based_2025_JFM}, as well as assessing coherence between sensor and target locations \citep{maia_real-time_2021,audiffred_reactive_2024}. The estimation error used throughout this paper is quantified by the metric \citep{martini_resolvent-based_2022,jung_toward_2024,jung_resovlent_based_2025_JFM}
\begin{equation}\label{eq_Errormetric}
E=\frac{\sum_{i} \int(\boldsymbol{\tilde{z}}_{i}(t)-\boldsymbol{z}_{i}(t))^{2} dt}{\sum_{i} \int(\boldsymbol{z}_{i}(t))^{2} dt},
\end{equation} 
where \(\boldsymbol{\tilde{z}}_{i}\) and \(\boldsymbol{z}_{i}\) represent the estimated and true values at the $i$-th target location, respectively. Additionally, coherence provides valuable insights into frequency-dependent estimation performance and is defined as \citep{towne_resolvent-based_2020}:
\begin{equation}\label{eq_Coherence}
\hat{\gamma}_{yz}(\omega)=\frac{|\mathsfbi{\hat{S}}_{yz}(\omega)|}{\sqrt{\mathsfbi{\hat{S}}_{yy}(\omega)}\sqrt{\mathsfbi{\hat{S}}_{zz}(\omega)}},
\end{equation}
where $\mathsfbi{\hat{S}}_{yz}$ denotes the cross-spectral density between sensor and target, and $\mathsfbi{\hat{S}}_{yy}$ and $\mathsfbi{\hat{S}}_{zz}$ denote their respective power spectral densities.\\
\indent Figure~\ref{Fig_EST_Error_field} presents spatial maps of estimation errors for a single sensor estimating streamwise velocity fluctuations at a single target location. Results are compared between the truncated non-causal (panels \textit{a}, \textit{c}, \textit{e}) and causal (panels \textit{b}, \textit{d}, \textit{f}) estimation approaches, covering three different flow representations: spanwise-averaged case (panels \textit{a}, \textit{b}), spanwise-Fourier modes at wavenumbers $k_{z}=10\pi$ (panel \textit{c}, \textit{d}), and mid-span plane case (panels \textit{e}, \textit{f}). For the spanwise-Fourier case, we focus specifically on the wavenumber $k_{z}=10\pi$, as it represents the lowest frequency and highest energy modes identified by previous resolvent gain studies \citep{yeh_resolvent-analysis-based_2019,towne_database_2023}.\\%
\indent Sensor locations are characterized using the coordinate $s$, defined such that positive values correspond to the suction surface and negative values to the pressure surface of the airfoil, with the magnitude representing the normalized streamwise position ($x/L_{c}$). Overall, the causal estimation method consistently outperforms the truncated non-causal approach. This advantage is particularly noticeable near the trailing-edge region ($1 < x/L_{c} < 1.5$) and prominently so when sensors are positioned on the front surfaces of the airfoil ($-0.5 < s < 0.5$). The performance improvement is more pronounced for the spanwise-averaged flow compared to the mid-span flow. This difference arises primarily due to the characteristics of the laminar region spanning $-0.5 < s < 0.5$ on the surface of the airfoil, where larger spanwise structures exist, resulting in less information about turbulent fluctuations in the wake, diminishing estimation accuracy. The detrimental impact of laminar regions on estimation accuracy is especially evident in the spanwise Fourier modes. For the spanwise Fourier modes, sensors located within the laminar region ($-0.5 < s < 0.5$), particularly on the pressure surface ($-0.5 < s < 0$), exhibit poor estimation performance. This effect is notably stronger at low wavenumbers, e.g., $k_{z}=10\pi$, likely due to its larger associated spanwise structures compared to higher wavenumbers, e.g., $k_{z}=20\pi$. However, downstream of the laminar-to-turbulent transition, estimation performance significantly improves due to the emergence of large-scale turbulent structures within the wake that are better correlated with the measurements.\\
\indent Figure~\ref{Fig_EST_generaltrend} presents the average estimation errors over the wake region as a function of sensor location $s$, comparing spanwise-averaged (S-A), spanwise-Fourier (S-F), and mid-span plane (M-S) flows. These results are shown separately for the streamwise velocity fluctuations ($u_{x}'$, panel \textit{a}) and the cross-streamwise velocity fluctuations ($u_{y}'$, panel \textit{b}). Here, we report the average error over a set of targets in the wake that are uniformly distributed along the line segment $1 < x/L_{c} < 2.5$ and $y/L_{c} = -0.11$.  The resolvent-based kernels, which inherently include the linearized Navier–Stokes operator, effectively estimate both velocity components from shear-stress sensor measurements. This effectiveness implies potential applicability of our method for estimating other relevant flow quantities, such as the vorticity field. Sensors placed on the pressure surface generally yield higher accuracy in spanwise-averaged and mid-span flows. Interestingly, the laminar region on the suction surface also performs well for estimation in spanwise-averaged flows, which contrasts with the results shown previously in figure~\ref{Fig_EST_Error_field}. This discrepancy arises because the targets analyzed in figure~\ref{Fig_EST_generaltrend} span both the upper and lower wake regions relative to the trailing edge, thereby incorporating regions of different coherence characteristics. Additionally, we observe a slight deterioration in estimation accuracy for sensors within the laminar separation bubble region ($0.4 < s < 0.7$), although accuracy improves progressively as the sensor location approaches the trailing edge. Overall, superior estimation performance is achieved for the spanwise-averaged flow compared to  the mid-span flow, with the spanwise Fourier modes exhibiting intermediate performance. Based on these observations, sensor positions near the trailing edge ($0.5 < x < 1$) and on the upstream suction ($ 0 < x < 0.3$) and pressure ($-1 < x < -0.7$) surfaces are identified as particularly effective locations for achieving high-performance resolvent-based estimation.\\
\begin{figure}[t]
  \begin{center}
      \begin{tikzpicture}[baseline]
        \tikzstyle{every node}=[font=\small]
        \tikzset{>=latex}
        \node[anchor=south west,inner sep=0] (image) at (0,0) {
          \includegraphics[scale=1,width=1\textwidth]{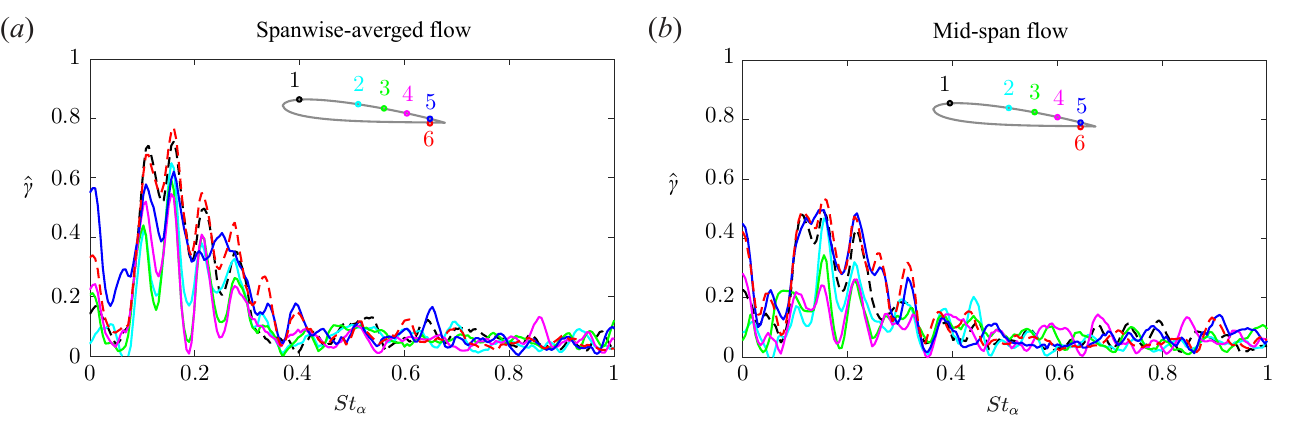}
        };
      \end{tikzpicture}
    \caption[]{\label{Fig_Coherence_K00_MID} Coherence of the streamwise velocity with respect to the target $z_{1} (x/L_{c}=1.2)$ for the spanwise-averaged and mid-span flows.}
  \end{center}
\end{figure}
\indent In figure \ref{Fig_Coherence_K00_MID}, we examine coherence of the streamwise velocity fluctuations between the sensor and the wake regions for the spanwise-averaged and mid-span flows at the sensor locations where single-sensor estimation was effective, as demonstrated in figures \ref{Fig_EST_Error_field} and \ref{Fig_EST_generaltrend}. These locations include regions on the front suction side ($0 < x/L_{c} < 0.3$), the rear pressure side ($-1 < x/L_{c} < -0.7$), and the mid-chord area ($0.5 < x/L_{c} < 1$), as illustrated in figures \ref{Fig_Coherence_K00_MID}. Since our goal in this coherence study is to gain physical insights into coherence between sensor locations and wake fluctuations, we focus our analysis on the near-wake target position at $x/L_{c}=1.2$. We confirmed that coherence trends at farther wake positions were similar but with reduced magnitudes. Specifically, the sensors representing the laminar regions were placed at 1 (suction side, laminar region $0 < x/L_{c} < 0.4$) and 6 (pressure side, laminar region $-1 < x/L_{c} < 0$), as shown in figure~\ref{Fig_Coherence_K00_MID}. Within each laminar region, the coherence patterns remain similar, differing only in magnitude. Notably, spanwise-averaged flow fields exhibit stronger coherence at frequencies of interest ($St_{\alpha}<0.2$) compared to mid-span flows. This enhanced coherence results from spatially larger coherent fluctuations associated with spanwise averaging, resulting in stronger coupling between sensor and target fluctuations. Interestingly, laminar regions (sensors 1 and 6) exhibit significant coherence with the near-wake target, suggesting that upstream laminar disturbances may be inherently linked to downstream wake fluctuations \citep{jung_resovlent_based_2025_JFM}. Furthermore, sensor 5, located in the turbulent region, demonstrates notable coherence with the target across a wide frequency range. Among middle locations (sensors 2, 3, and 4), each sensor displays pronounced coherence at the frequency bands of interest ($St_{\alpha}<0.2$). These locations are empirically selected to capture dominant frequencies at the target, which are crucial for kernel-based amplification in the frequency domain.\\ 
\begin{figure}[t]
  \begin{center}
      \begin{tikzpicture}[baseline]
        \tikzstyle{every node}=[font=\small]
        \tikzset{>=latex}
        \node[anchor=south west,inner sep=0] (image) at (0,0) {
          \includegraphics[scale=0.6,width=0.6\textwidth]{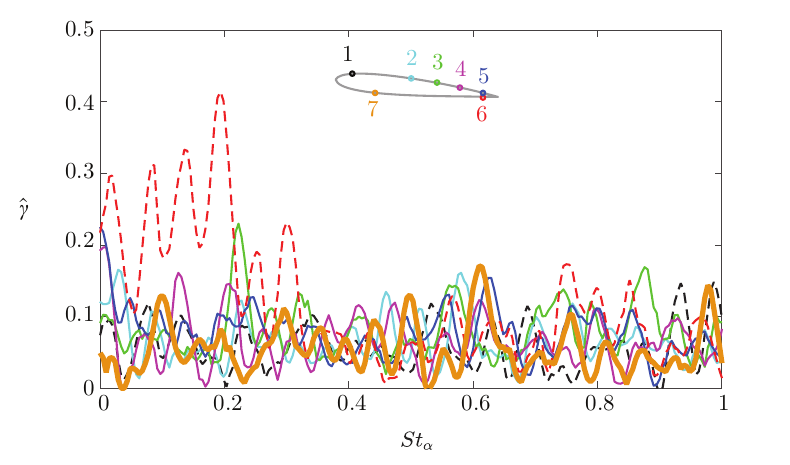}
        };
      \end{tikzpicture}
    \caption[]{\label{Fig_Coherence_K05} Coherence of the streamwise velocity with respect to the target $z_{1} (x/L_{c}=1.2)$ of the spanwise-Fourier case at $k_{z}=10\pi$.}
  \end{center}
\end{figure}
\indent We further investigated coherence between the sensor and the wake regions with respect to the streamwise velocity fluctuations for the spanwise-Fourier mode at $k_{z}=10\pi$. In particular, we are interested in verifying whether coherence is indeed weak in the region $-0.7 < x/L_{c} < 0$, as shown in the single-sensor estimation result in figure~\ref{Fig_EST_Error_field}. To assess this, we examined coherence at sensor location~7, placed at $x/L_{c} = -0.3$, as indicated by the yellow solid line in figure~\ref{Fig_Coherence_K05}. Although the von Kármán vortex shedding and other dominant frequencies are evident for $St_{\alpha}<0.2$, sensor~7 does not exhibit significant coherence in this frequency range. In contrast, sensor locations~4 (purple), 5 (blue), and 6 (red) display prominent coherence peaks, consistent with effective estimation performance. Figure \ref{Fig_Coherence_K05} further indicates that sensors 5 and 6 exhibit increased coherence at a spanwise frequency of $10\pi$. Ultimately, we identify sensors 1, 5, and 6 as the most effective for estimation purposes, with an additional selection of sensor 4 from among sensors 2, 3, and 4.\\
\subsection{Power-spectral density of the sensor and target readings}\label{subsec_PSD}

\begin{figure}[t]
  \begin{center}
      \begin{tikzpicture}[baseline]
        \tikzstyle{every node}=[font=\small]
        \tikzset{>=latex}
        \node[anchor=south west,inner sep=0] (image) at (0,0) {
          \includegraphics[scale=0.9,width=0.9\textwidth]{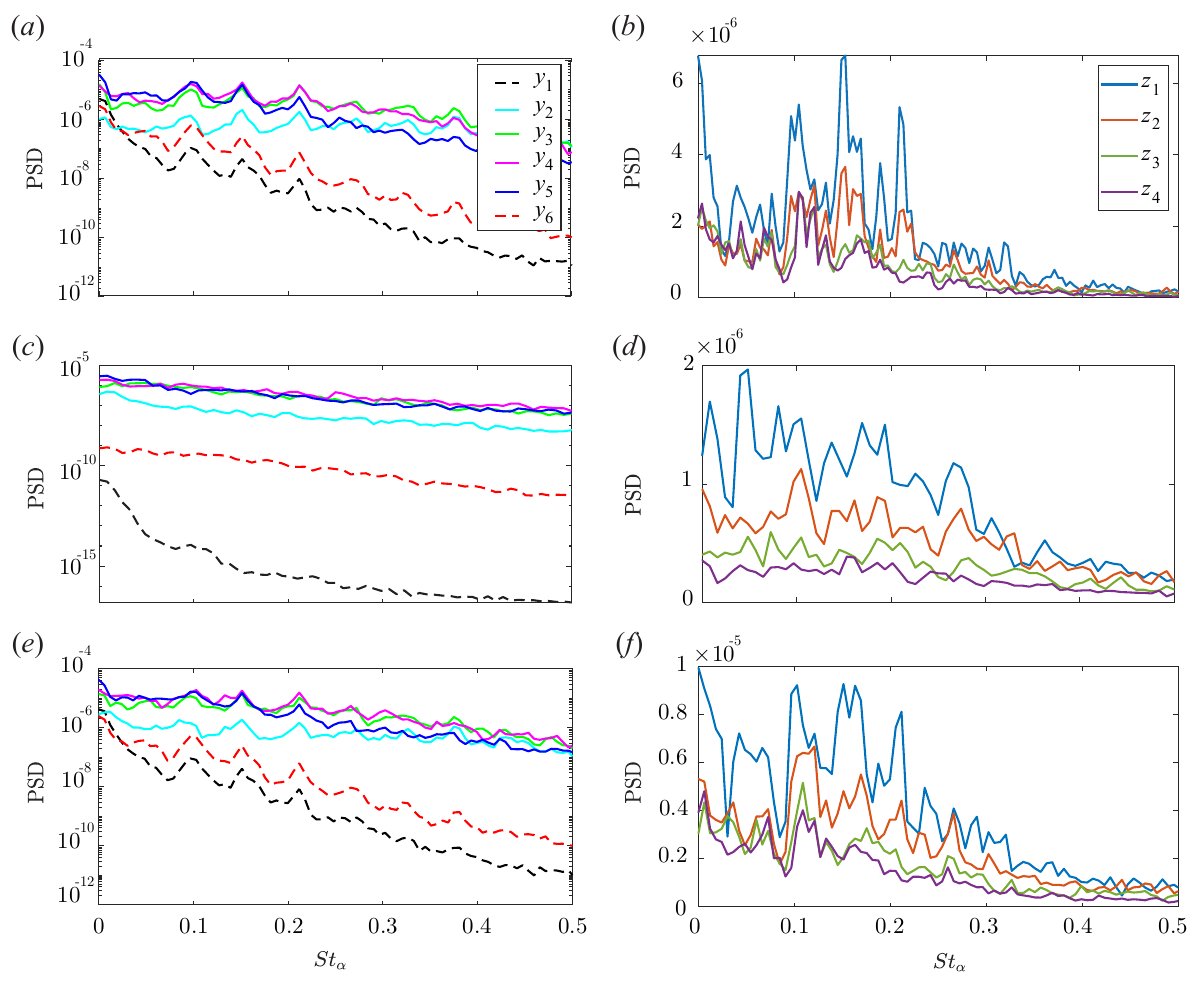}
        };
      \end{tikzpicture}
    \caption[]{\label{Fig_PSDyy_zz} PSD of streamwise velocity fluctuations $u_{x}'$ for sensor readings and target readings, obtained from the spanwise-averaged (\textit{a}) and (\textit{b}), spanwise-Fourier ($k_z=10\pi$) (\textit{c}) and (\textit{d}), and mid-span plane (\textit{e}) and (\textit{f}) cases presented in figure \ref{Fig_velfield_all}. In (\textit{a}), the dashed lines indicate positions $y_1$ and $y_6$, corresponding to the laminar flow region.}
  \end{center}
\end{figure}
\indent Figure~\ref{Fig_PSDyy_zz} presents the power spectral density (PSD) of the sensor and target signals for the spanwise-averaged, spanwise-Fourier, and mid-span plane flows. Panels (\textit{a}) and (\textit{b}) highlight three dominant frequencies evident in the signals at sensor positions $y_{3}$, $y_{4}$, and $y_{5}$, as well as the target position $z_{1}$. The energy associated with these dominant frequencies decreases downstream, as demonstrated by the PSD at target positions $z_2$, $z_3$, and $z_4$ in panel (\textit{b}). It is well established that laminar separation bubble (LSB) detachment over the suction surface promotes the Kelvin--Helmholtz instability, resulting in vortex shedding characterized by lower-frequency oscillations ($St_{\alpha}<0.12$) \citep{ducoin_numerical_2016}. These lower-frequency peaks are clearly captured in panel (\textit{a}) for sensor positions $y_{3}$, $y_{4}$, $y_{5}$, and the near-wake target $z_{1}$. In contrast, the turbulent region on the airfoil surface and near wake exhibit dominant von Kármán vortex shedding frequencies within the higher-frequency band ($0.152 \leq St_{\alpha} \leq 0.211$) \citep{yeh_resolvent-analysis-based_2019}.\\
\indent Figure~\ref{Fig_PSDyy_zz}(\textit{c}) and (\textit{d}) illustrate the PSD of sensor and target measurements for the spanwise-Fourier mode at the wavenumber $k_{z}=10\pi$, respectively. The PSD computations utilize the same parameters as those used for the spanwise-averaged case. Compared to the spanwise-averaged (\textit{a}) and mid-span plane (\textit{e}) cases, the PSD at the sensor and target locations (\textit{c}) does not exhibit distinct dominant peaks, consistent with the coherence results shown in figure~\ref{Fig_Coherence_K05}. As observed in \citet{yeh_resolvent-analysis-based_2019} and \citet{towne_database_2023}, the response modes for spanwise-Fourier modes at $k_{z}=10\pi$ appear predominantly slightly away from the airfoil surface. Thus, the spanwise Fourier modes at the sensor locations are relatively weak, leading to weaker PSD values, as illustrated in figure~\ref{Fig_PSDyy_zz}(\textit{c}). Modes at lower wavenumbers correspond to larger, more energetic spanwise structures, whereas modes at higher wavenumbers represent smaller, less energetic structures. We confirmed that the primary difference between the wavenumbers $k_z=10\pi$ and $20\pi$ is the magnitude of their PSD values, while their PSD patterns remain similar.\\
\indent Figure~\ref{Fig_PSDyy_zz}(\textit{e}) and (\textit{f}) display the PSD of sensor and target signals in the mid-span plane. Similar to the spanwise-averaged case, the laminar flow region ($y_{1}$ and $y_{6}$) exhibits lower energy fluctuations. Additionally, energy progressively dissipates as the target moves further downstream. Dominant frequencies in the approximate range $0.1 < St_{\alpha} < 0.2$ are distinctly visible in the near-wake targets ($z_{1}$ and $z_{2}$), as clearly illustrated in panel (\textit{f}).
\subsection{Estimation kernels}\label{subsec_Kernels}
\begin{figure}[t]
  \begin{center}
      \begin{tikzpicture}[baseline]
        \tikzstyle{every node}=[font=\small]
        \tikzset{>=latex}
        \node[anchor=south west,inner sep=0] (image) at (0,0) {
          \includegraphics[scale=1,width=0.80\textwidth]{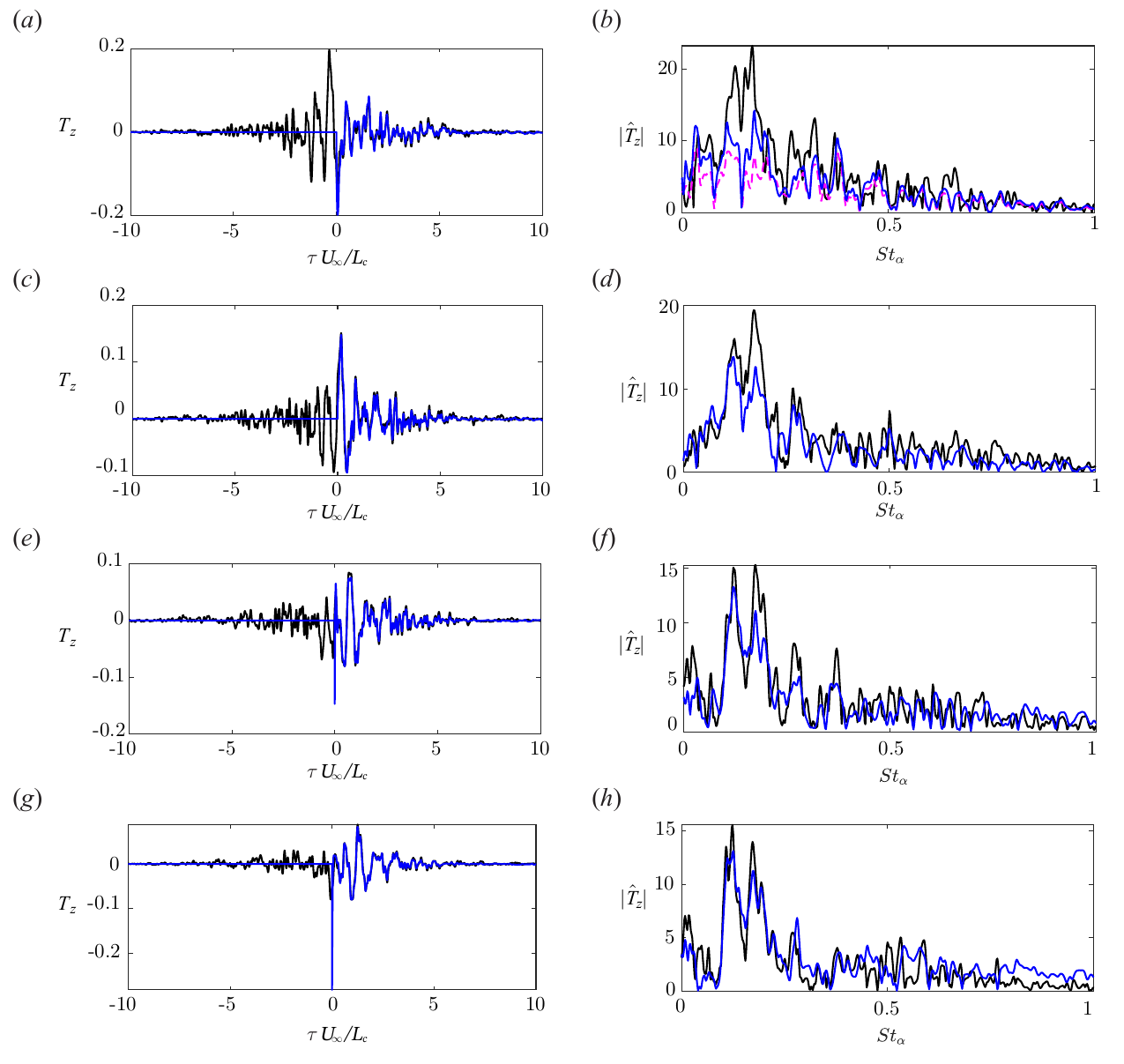}
        };
      \end{tikzpicture}
    \caption[]{\label{Fig_TF_spanwiseAVG1} Estimation kernels between sensor $y_{1}$ and targets $z_{1}$ (\textit{a}, \textit{b}), $z_{2}$ (\textit{c}, \textit{d}), $z_{3}$ (\textit{e}, \textit{f}), and $z_{4}$ (\textit{g}, \textit{h}) for the spanwise-averaged flow. Kernels are shown in the time domain (left column) and frequency domain (right column). The black line represents the non-causal kernel, the blue line represents the causal kernel obtained via Wiener--Hopf decomposition, and the magenta dashed line (when shown) indicates the truncated non-causal kernel.}
  \end{center}
\end{figure}
\begin{figure}[t]
  \begin{center}
      \begin{tikzpicture}[baseline]
        \tikzstyle{every node}=[font=\small]
        \tikzset{>=latex}
        \node[anchor=south west,inner sep=0] (image) at (0,0) {
          \includegraphics[scale=1,width=0.8\textwidth]{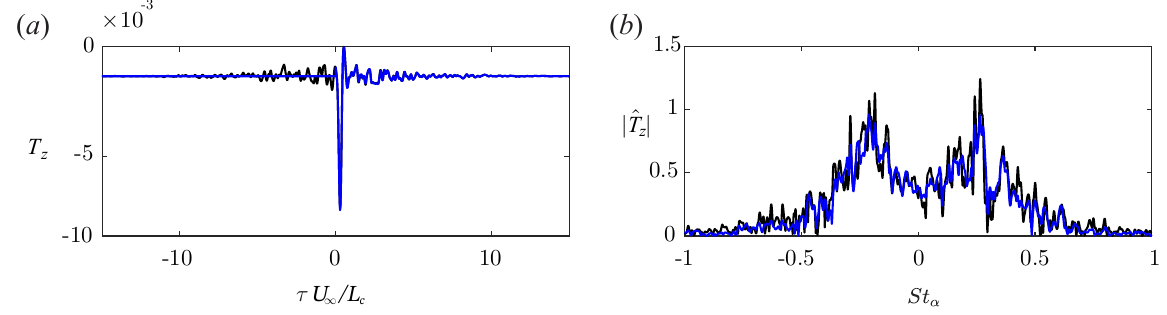}
        };
      \end{tikzpicture}
    \caption[]{\label{Fig_ESTKERNEL_KZ05} Estimation kernels between sensor $y_{5}$ and target $z_{1}$ for the spanwise Fourier mode at $k_{z}=10\pi$, shown in the time (\textit{a}) and frequency (\textit{b}) domains. Sensor and target positions are indicated in figure~\ref{Fig_velfield_all}. The black line represents the non-causal kernel, and the blue line represents the causal kernel obtained via Wiener--Hopf decomposition.}
  \end{center}
\end{figure}
Next, we examine the estimation kernels, building on our previous studies \citep{martini_resolvent-based_2020, martini_resolvent-based_2022, jung_resovlent_based_2025_JFM}, where we initially explored kernels within the resolvent-based framework. In this paper, we derive the kernels using a data-driven approach that accounts for the statistical effects of flow nonlinearity, implicitly incorporating the CSD matrix $\hat{\mathsfbi{F}}_{nl}$, which acts as a forcing in linear dynamics. Note that when white noise forcing is applied, the estimation kernels correspond to the transfer functions of the Kalman filter \citep{Kalman_1960}, as shown by \citet{martini_resolvent-based_2022}. Incorporating the forcing matrix $\hat{\mathsfbi{F}}_{nl}$ is suitable for this study, as our application involves strongly nonlinear turbulent flows. Modeling the nonlinear terms for estimation leads to better performance. A key characteristic of estimation kernels is the presence of peak points, which reveal the time it takes for hydrodynamic waves to travel from the sensor location to the target. Since the estimation kernels are based on the compressible linearized Navier-Stokes operator, they also capture acoustic waves, resulting in secondary peaks in addition to the hydrodynamic waves \citep{jung_resolvent-based_2023, towne_resolvent-based_2024}. Another important characteristic is whether there are significant values in the non-causal part of the kernel ($\tau<0$), including the peak point \citep{Jung_RSV}. This information is crucial for estimation, but it may be truncated when the estimator is applied to the convolution function if it resides in the non-causal part of the kernels. This truncation can lead to a significant difference in the performance between the causal and truncated non-causal approaches. From a numerical perspective, the estimation kernels function by amplifying the frequency signals of the sensor measurements through multiplication with the kernels to approximate the target signals, i.e., $\hat{\tilde{\boldsymbol{z}}}(\omega)=\hat{\mathsfbi{T}}_{z}(\omega)\hat{\boldsymbol{y}}(\omega)$. Therefore, analyzing the estimation kernels in the frequency domain alongside the PSD of the sensor and target readings in $\S$\ref{subsec_PSD} provides valuable numerical insights into accurate estimation.\\
\indent We examine four kernels, specifically within the single input and multiple outputs configuration (sensor: $y_{1}$ and targets: $z_{1}$, $z_{2}$, $z_{3}$, $z_{4}$), as illustrated in figure \ref{Fig_TF_spanwiseAVG1}. This selection is based on the observation that when the sensor is positioned in the laminar region within the effective locations, the kernel's peaks are distinct, with less noise in other frequencies, showing a clear correlation between the sensor and the target signals. In figure \ref{Fig_TF_spanwiseAVG1}, prominent peak points are observed in the non-causal kernels (\textit{a}, \textit{c}, \textit{e}, \textit{g}) at [$\tau_{\text{peak},z_{1}}^{},\tau_{\text{peak},z_{2}}^{},\tau_{\text{peak},z_{3}}^{},\tau_{\text{peak},z_{4}}^{}] = [-0.360, 0.171, 0.676, 1.25]$, respectively. These points represent the primary hydrodynamic wave travel times between the sensor and the targets. The case in figure \ref{Fig_TF_spanwiseAVG1}(\textit{a}) serves as a good example of how a causal approach using Wiener-Hopf decomposition can enhance estimation accuracy. When this kernel is applied to real-time estimation for evaluating the convolution function, the peak at $\tau=-0.36$, which contains critical travel information of the fluctuations, is truncated. The truncated non-causal kernel in the frequency domain (magenta dashed line) is notably lower in magnitude compared to the causal kernel (blue line), as shown in \ref{Fig_TF_spanwiseAVG1}(\textit{b}). It is important to note that the kernels amplify the sensor measurements at dominant frequencies ($0.1 < St_{\alpha} < 0.2$).\\
\indent Imposing sensor is necessary to ensure that the inversion of the PSD is well-posed \citep{martini_resolvent-based_2020,martini_resolvent-based_2022}. The noise level, which is used to regularize the kernels, is set at $10^{-2}$ of the maximum PSD value of the sensor measurements, consistent with the approach used in previous work \citep{martini_resolvent-based_2022, jung_resovlent_based_2025_JFM}. Similar to other optimal estimation and control methods \citep{chevalier_state_2006, schmid_linear_2016}, the sensor noise is described by correlations that are assumed to be known.\\
\begin{figure}[t]
  \begin{center}
      \begin{tikzpicture}[baseline]
        \tikzstyle{every node}=[font=\small]
        \tikzset{>=latex}
        \node[anchor=south west,inner sep=0] (image) at (0,0) {
          \includegraphics[scale=1,width=0.8\textwidth]{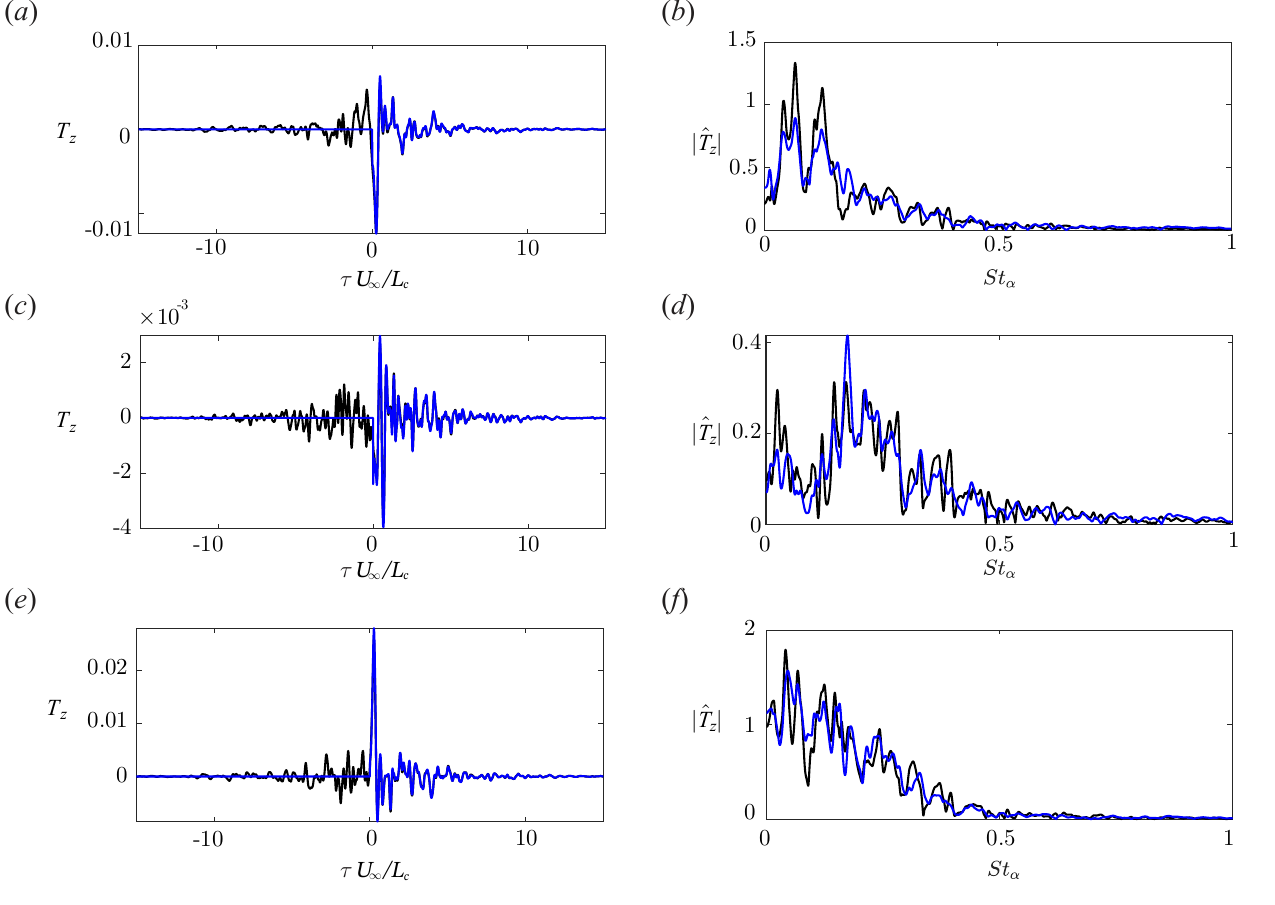}
        };
      \end{tikzpicture}
    \caption[]{\label{Fig_est_midspan_Tz} Estimation kernels between sensors $y_{1}$ (\textit{a}, \textit{b}), $y_{5}$ (\textit{c}, \textit{d}), and $y_{6}$ (\textit{e}, \textit{f}) and target $z_{1}$ for the mid-span plane case. Kernels are shown in the time domain (left column) and frequency domain (right column). The black line represents the non-causal kernel, and the blue line represents the causal kernel obtained via Wiener--Hopf decomposition.}
  \end{center}
\end{figure}
\indent Figure \ref{Fig_ESTKERNEL_KZ05} shows the kernels for the spanwise-Fourier modes at $k_{z}=10\pi$. Based on figure~\ref{Fig_EST_Error_field}, sensor placement in the laminar region is less effective; hence, we focus on kernels corresponding to sensor $y_{5}$, rather than $y_{1}$, and target $z_{1}$, consistent with locations examined previously in figure~\ref{Fig_TF_spanwiseAVG1}. Because the spanwise Fourier-transformed state is complex-valued, the computational cost for estimation increases by a factor of four. Additionally, the kernels are no longer symmetric at $St_{\alpha}=0$, as shown in figure \ref{Fig_ESTKERNEL_KZ05}. The peak points are clearly visible (see figures \ref{Fig_ESTKERNEL_KZ05}(\textit{a}) and (\textit{c})) at $\tau > 0$, indicating the dominance of hydrodynamic waves. Consequently, performance differences between causal and truncated non-causal estimation methods are minimal. The values of the kernels are largest in the range of $|St_{\alpha}|<0.2$, effectively amplifying sensor measurements to produce accurate estimates.\\
\indent Finally, we present the kernels for the mid-span plane flow in figure~\ref{Fig_est_midspan_Tz}. Sensors positioned in both the 
laminar ($y_{1}$) and turbulent regions (suction side: $y_{5}$; pressure side:
$y_{6}$) are shown. The prominent peak points in the kernels between sensors ($y_{1}$, $y_{5}$) and target $z_{1}$ closely match those observed in the spanwise-averaged and spanwise-Fourier flows, respectively, as depicted in figures \ref{Fig_TF_spanwiseAVG1} and \ref{Fig_ESTKERNEL_KZ05} for the kernels between $y_1$ and $z_1$ and $y_5$ and $z_1$, respectively. The kernel between $y_{6}$ and $z_{1}$ also aligns well with the other cases, though it is not explicitly shown here. However, the kernels for the mid-span flow exhibit significant amplitudes over a larger range of $\tau$, indicating greater difficulty in achieving temporal convergence. A notable difference appears in figure~\ref{Fig_est_midspan_Tz}(\textit{a}), where the kernel peak lies in the positive domain ($\tau>0$), unlike the corresponding kernel in the spanwise-averaged flow, which peaks in the negative domain. This discrepancy may be attributed to the fact that the spanwise-averaged case integrates the flow across multiple spanwise positions, capturing coherent large-scale structures that extend significantly in the spanwise direction. These large-scale structures are more capable of influencing upstream conditions (non-causal behavior), appearing as peaks in the non-causal part ($\tau<0$) of the kernel. In contrast, the mid-span plane case represents a single spanwise slice, emphasizing smaller-scale, localized spatial structures. Such structures typically have limited upstream influence since their acoustic disturbances are weaker, more rapidly dissipated, and less coherent over larger spanwise distances. Thus, such conditions enhance convective flow characteristics in the mid-span plane case, thereby improving the accuracy when applying truncated non-causal estimation to the mid-span plane flow. Additionally, we observe that the kernel between $y_{4}$ and $z_{1}$ is similar to that between $y_{5}$ and $z_{1}$ within the mid-span flow, so it is omitted from this figure. Both causal and truncated non-causal kernels for targets further downstream ($z_{2}$, $z_{3}$, $z_{4}$) closely resemble those in the spanwise-averaged flow (figure~\ref{Fig_TF_spanwiseAVG1}), with their peak points consistently located in the positive $\tau$ domain.\\
\subsection{Multi-sensor estimation}\label{subsec_Estimation_results}
\begin{figure}[t]
  \begin{center}
      \begin{tikzpicture}[baseline]
        \tikzstyle{every node}=[font=\small]
        \tikzset{>=latex}
        \node[anchor=south west,inner sep=0] (image) at (0,0) {
          \includegraphics[scale=1,width=1\textwidth]{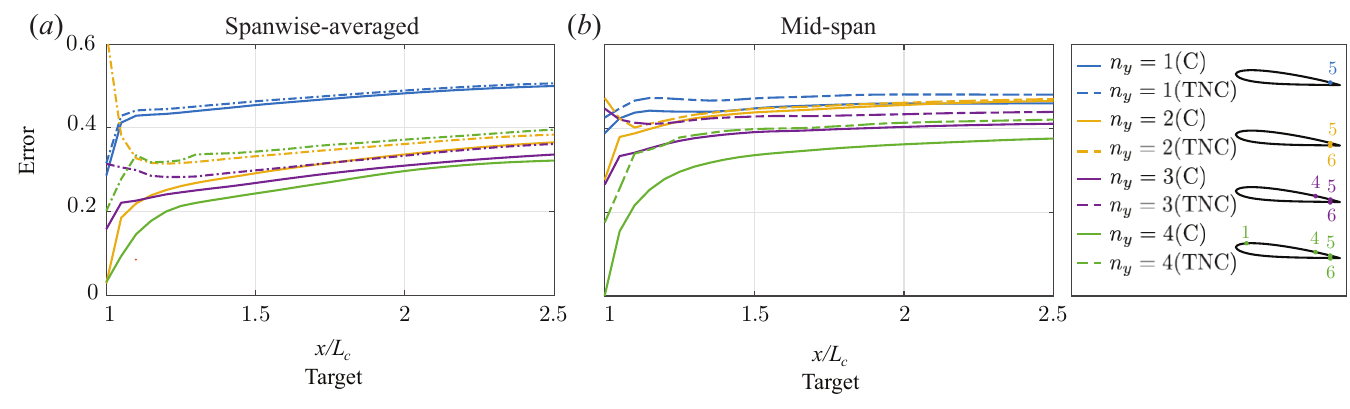}
        };
      \end{tikzpicture}
    \caption[]{\label{Fig_EST_error_downstream_increasing_sensor} Estimation error along the trailing edge for spanwise-averaged and mid-span flows using one sensor ($y_{5}$), two sensors ($y_{5}$, $y_{6}$), three sensors ($y_{4}$, $y_{5}$, $y_{6}$), and four sensors ($y_{1}$, $y_{4}$, $y_{5}$, $y_{6}$).}
  \end{center}
\end{figure}
\begin{figure}[t]
  \begin{center}
      \begin{tikzpicture}[baseline]
        \tikzstyle{every node}=[font=\small]
        \tikzset{>=latex}
        \node[anchor=south west,inner sep=0] (image) at (0,0) {
          \includegraphics[scale=0.8,width=0.8\textwidth]{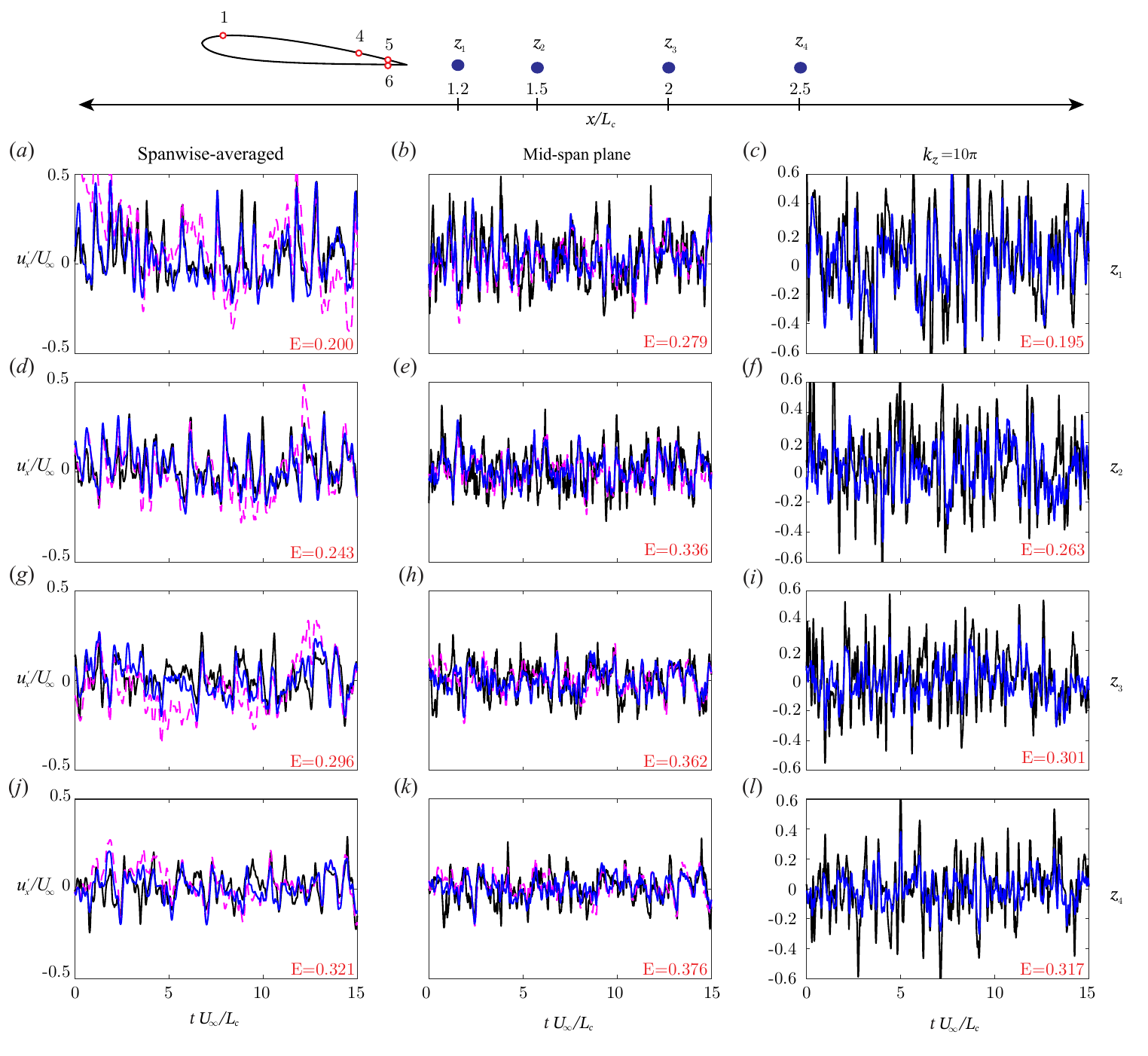}
        };
      \end{tikzpicture}
    \caption[]{\label{Fig_EST_ALL_4sensors} Estimation of $u_x'$ as a function of time using four sensors ($y_1$, $y_4$, $y_5$, $y_6$): spanwise-averaged (panels \textit{a, d, g, j}), mid-span plane (panels \textit{b, e, h, k}), spanwise-Fourier cases at $k_z = 10\pi$ (panels \textit{c, f, i, l}). Target locations are defined by coordinates $[z_1 = x/L_c,\; y/L_c]$: panels (\textit{a}--\textit{c}) at $[1.2, -0.11]$, (\textit{d}--\textit{f}) at $[1.5, -0.11]$, (\textit{g}--\textit{i}) at $[2.0, -0.11]$, and (\textit{j}--\textit{l}) at $[2.5, -0.11]$, as illustrated in the top figure. The true data from LES are shown in black, causal estimations in blue, and truncated non-causal estimations as magenta dashed lines. The causal estimation error is indicated in red in the bottom right corner of each panel.}
  \end{center}
\end{figure}
\begin{figure}[t]
  \begin{center}
      \begin{tikzpicture}[baseline]
        \tikzstyle{every node}=[font=\small]
        \tikzset{>=latex}
        \node[anchor=south west,inner sep=0] (image) at (0,0) {
          \includegraphics[scale=1,width=1\textwidth]{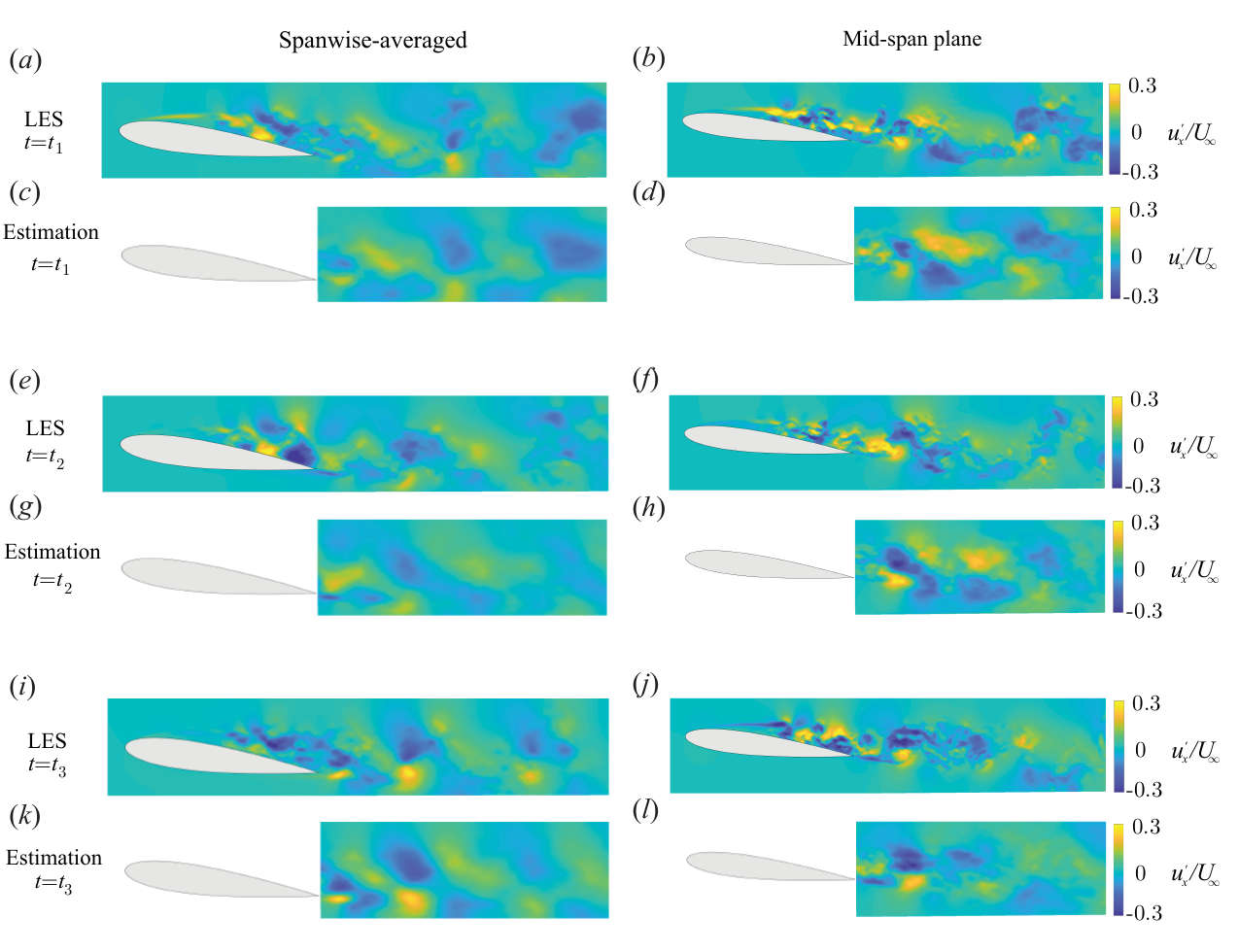}
        };
      \end{tikzpicture}
    \caption[]{\label{Fig_EST_highrank_ALL} LES and estimated snapshots of $u_{x}'$ across the extended wake region for the spanwise-averaged (panels \textit{a, c, e, g, i, k})  and mid-span plane (panels \textit{b, d, f, h, j, l}) flows, utilizing four sensors ($y_{1},y_{4},y_{5},y_{6}$). The times $t_1$, $t_2$, and $t_3$ were chosen to represent three distinct phases based on the dominant vortex shedding frequency.}
  \end{center}
\end{figure}
\indent By selecting sensor locations $y_{1}$, $y_{4}$, $y_{5}$, and $y_{6}$ (notations from figure \ref{Fig_Coherence_K00_MID}(\textit{a}) and (\textit{b})) based on the investigation of the most effective locations using single-sensor estimation and coherence in $\S$\ref{sub_sensorplacement}, we show the estimation errors for spanwise-averaged and mid-span flows along the trailing edge as additional sensors are introduced, as shown in figure \ref{Fig_EST_error_downstream_increasing_sensor}. The use of sensor $y_{5}$ (blue) alone does not yield significant improvement over truncated non-causal estimation. However, adding an additional sensor ($y_{6}$) markedly enhances estimation accuracy (yellow). With these two sensors, truncated non-causal estimation performs poorly near the trailing edge, whereas causal estimation remains accurate. The inclusion of a third sensor $y_{4}$ (purple) alongside $y_{5}$ and $y_{6}$ results in a slight improvement. The best estimation results are achieved using all four sensors $y_{1}$, $y_{4}$, $y_{5}$, $y_{6}$ (green). These findings demonstrate greater accuracy than the use of six sensors uniformly distributed on the pressure side of the airfoil, as reported in our previous work \citep{jung_toward_2024}.\\
\indent Figure~\ref{Fig_EST_ALL_4sensors} shows the time-series estimation of streamwise velocity fluctuations using four sensors (\(y_{1}, y_{4}, y_{5}, y_{6}\)) for spanwise-averaged (panels \textit{a, d, g, j}), mid-span plane (panels \textit{b, e, h, k}), and spanwise-Fourier (panels \textit{c, f, i, l}) cases. The causal estimation error, computed using \eqref{eq_Errormetric}, is indicated in the bottom right corner of each panel. As expected, the estimation error increases with greater distance between the sensor and the target location. The Wiener–Hopf approach significantly improves estimation accuracy for targets located closer to the trailing edge, as evident in figure~\ref{Fig_EST_ALL_4sensors}(\textit{a}). The difference in accuracy between truncated non-causal (TNC) and causal methods diminishes as the target moves further downstream. Among the investigated flow types, the spanwise-averaged flow, characterized by less chaotic spatial structures, is estimated most accurately. For the mid-span plane flow, presented in figure~\ref{Fig_EST_ALL_4sensors} (panels~\textit{b, e, h, k}), the time-series velocity fluctuations exhibit more random oscillations compared to the other flow types considered. However, despite this increased complexity, the broader trends in the velocity fluctuations over time are still effectively captured. Although the sensor may not fully capture the detailed footprint of turbulence, the estimation remains effective. These results are consistent with the trends shown in figure~\ref{Fig_EST_error_downstream_increasing_sensor}.\\
\indent Figure~\ref{Fig_EST_ALL_4sensors} (panels~\textit{c, f, i, l}) displays time-series estimations of the real part of \(u_{x}'\) for the spanwise-Fourier modes with \(k_{z}=10\pi\). As anticipated, estimation accuracy decreases with increasing spanwise wavenumber. Kernels derived from resolvent operators at frequencies associated with lower energetic gains provide lower estimation accuracy. Since the TNC and causal estimations for the spanwise-Fourier modes yield similar results, the TNC estimations are not shown explicitly in figure~\ref{Fig_EST_ALL_4sensors} for these cases.\\
\indent Figure~\ref{Fig_EST_highrank_ALL} shows three random snapshots of the causal estimation in the extended wake region for spanwise-averaged and mid-span plane cases, using the same sensor configuration as in figure~\ref{Fig_EST_ALL_4sensors}. The estimation errors for the wake region, calculated using metric \ref{eq_Errormetric}, are approximately 0.254 and 0.324 for causal estimation, and 0.352 and 0.389 for truncated non-causal estimation in the spanwise-averaged and mid-span plane cases, respectively. Although the small-scale structures are not fully captured due to the limitations of the sensors in detecting turbulence, the causal resolvent-based estimator effectively estimates the wake fluctuations. Given that the estimation relies solely on a limited number (four sensors) of real-time measurements, these results demonstrate significant promise for practical applications in experimental settings and other engineering scenarios.

\section{Conclusions}\label{conclusion}

\indent In this study, we developed and demonstrated a causal resolvent-based framework for estimating turbulent velocity fluctuations in the wake of a spanwise-periodic NACA0012 airfoil at moderate Reynolds $\emph{Re}_{L{c}} = 23,000$, $Ma_{\infty} \equiv U_{\infty}/c_{\infty} = 0.3$, and an angle of attack of $\alpha = 6^\circ$. Building on previous research focused primarily on a laminar and globally stable flow scenario \citep{jung_resovlent_based_2025_JFM}, our approach extended resolvent-based estimation into the more complex regime of a turbulent, globally unstable wake flow.\\
\indent To address the key challenges inherent in this turbulent regime, including global instabilities, broadband multiscale dynamics, and large datasets, we employed a data-driven approach. This allowed construction of resolvent-based kernels from cross-spectral densities directly obtained via large-eddy simulations, inherently incorporating the colored-in-time statistical properties of nonlinear forcing terms. A key improvement was the adoption of the Wiener–Hopf method to enforce causality \citep{martini_resolvent-based_2022}, which significantly enhanced real-time estimation accuracy compared to traditional truncated, non-causal estimators. To handle computational complexity arising from high-dimensional datasets, our methodology utilized scalable parallel algorithms integrated within an existing compressible flow solver \citep{jung_resolvent-based_2024, towne_resolvent-based_2024}.\\
\indent Our results illustrated successful causal estimation across spanwise-averaged, spanwise-Fourier, and mid-span plane representations of the turbulent wake, using only limited shear-stress measurements from the airfoil surface. By strategically analyzing single-sensor estimation errors and coherence between sensor and target locations, we identified practical sensor-placement strategies that significantly enhanced estimation fidelity. Ultimately, our approach provided accurate, interpretable, and scalable predictions of an extended portion of the wake, capturing dominant coherent structures crucial for both scientific understanding and potential flow control applications.\\
\indent Future research directions suggested by this study include developing formal optimization methods for sensor placement and incorporating three-dimensional sensor configurations to further enhance estimation accuracy. The success of the resolvent-based estimation framework suggests that resolvent-based closed-loop flow control strategies could help mitigate turbulent fluctuations, reduce drag, increase lift, and minimize aerodynamic noise. It would also be valuable to investigate the generalizability and robustness of this approach under varying flow conditions, different airfoil geometries, higher Reynolds numbers, and more complex aerodynamic environments. Integrating advanced machine-learning techniques, such as neural networks, into the resolvent-based framework would further improve real-time computational efficiency and effectively handle larger datasets. Finally, experimental validation of the proposed resolvent-based estimation approach in practical wind-tunnel tests or flight scenarios is recommended to confirm its real-world applicability and robustness. By pursuing these avenues, our resolvent-based estimation framework has the potential to significantly advance the predictive modeling, real-time monitoring, and control of complex turbulent aerodynamic flows.
\section*{Funding}
This research was supported by the Air Force Office of Scientific Research (AFOSR) under grant number FA9550-20-1-0214. Computational resources were provided by the U.S. Department of Defense High-Performance Computing Modernization Program.

\section*{Declaration of interests}
The authors report no conflict of interest.
\section*{Author ORCIDs}
\orcidlink{https://orcid.org} Junoh Jung \href{https://orcid.org/0000-0003-0962-3127}{https://orcid.org/0000-0003-0962-3127};\\
\orcidlink{https://orcid.org} Aaron Towne \href{https://orcid.org/0000-0002-7315-5375}{https://orcid.org/0000-0002-7315-5375}.

\begin{appendices}

\section{Wiener-Hopf method}\label{app_WH}
We briefly review the theoretical and numerical Wiener–Hopf decompositions employed for enforcing causality within the resolvent-based estimation and control framework described by \citet{martini_resolvent-based_2022, audiffred_experimental_2023,jung_resovlent_based_2025_JFM}.
\subsection{Theoretical Wiener-Hopf decomposition}

The Wiener-Hopf method \citep{noble1958} decomposes frequency-domain functions into causal (analytic in lower half-plane, $(+)$) and non-causal (analytic in upper half-plane, $(-)$) components. This decomposition is employed here to enforce causality on estimation and control kernels, following \citet{daniele_fredholm_2007,martinelli_feedback_nodate}.

We consider the Wiener-Hopf problem
\begin{equation}
\mathsfbi{\hat{H}}(\omega)\mathsfbi{\hat{\Gamma}}_{+}(\omega) = \mathsfbi{\hat{\Lambda}}_{-}(\omega)+\mathsfbi{\hat{G}}(\omega),
\end{equation}
where $\mathsfbi{\hat{H}}$ and $\mathsfbi{\hat{G}}$ are known matrices, while $\mathsfbi{\hat{\Gamma}}_{+}$ and $\mathsfbi{\hat{\Lambda}}_{-}$ are unknown.

The Wiener-Hopf method involves additive and multiplicative factorizations, defined as
\begin{equation}
\mathsfbi{\hat{\Gamma}}(\omega)=[\mathsfbi{\hat{\Gamma}}(\omega)]_{-}+[\mathsfbi{\hat{\Gamma}}(\omega)]_{+}, \quad
\mathsfbi{\hat{\Gamma}}(\omega)=(\mathsfbi{\hat{\Gamma}}(\omega))_{-}(\mathsfbi{\hat{\Gamma}}(\omega))_{+},
\end{equation}
respectively. The explicit solution for the above problem is
\begin{equation}
\mathsfbi{\hat{\Gamma}}_{+}(\omega)=\left[\mathsfbi{\hat{G}}(\omega)(\mathsfbi{\hat{H}}(\omega))_{-}^{-1}\right]_{+}(\mathsfbi{\hat{H}}(\omega))_{+}^{-1}.
\end{equation}

\subsection{Numerical Wiener-Hopf decomposition}

Numerically, the additive factorization is straightforward, while the multiplicative factorization requires solving an integral equation. Following \citet{martini_resolvent-based_2022}, this integral equation is given by
\begin{equation}
\hat{\boldsymbol{x}}_i(\omega)+\frac{1}{2 \mathrm{i}}\mathcal{H}(\hat{\boldsymbol{x}}_i)(\omega)-\frac{1}{2 \mathrm{i}}\mathsfbi{\hat{G}}^{-1}(\omega)\mathcal{H}(\mathsfbi{\hat{G}}\hat{\boldsymbol{x}}_i)(\omega)=\mathsfbi{\hat{G}}^{-1}(\omega)\frac{\hat{\boldsymbol{w}}_{i,-}(\omega_0)}{\omega-\omega_0},
\end{equation}
where $\mathcal{H}$ denotes the Hilbert transform defined by
\begin{equation}
\mathcal{H}(\hat{\boldsymbol{x}})=P.V.\frac{1}{\pi}\int_{-\infty}^{\infty}\frac{\hat{\boldsymbol{x}}(u)}{\omega-u}\,\mathrm{d}u.
\end{equation}

The reader is referred to \citet{noble1958,daniele_fredholm_2007,martinelli_feedback_nodate,martini_resolvent-based_2022} for a detailed discussion of the Wiener–Hopf method. We numerically solve this integral equation using the GMRES iterative method, implemented directly within the CFD solver. This approach ensures computational efficiency and minimizes the need for external post-processing \citep{jung_resovlent_based_2025_JFM}.

 \section{Convergence of the data-driven estimation kernels}\label{app_Conv_kernels}
\begin{table}[h]
 \centering
\begin{tabular}{ c c c c c c}
\hline

Case & $N_{freq}$ & $N_{ovlp}$ &  $N_{blks}$& Error\\
\hline
1 & 500 &  250 & 296  &  0.400748 	\\
2 & 1,000 &  500 & 148 &  0.350644	\\
3 & 2,000 &  1,000 & 74 &  0.330292 	\\
4 & 3,000 &  1,500 & 48 &  0.320681 	\\
5 & 4,000 &  2,000 & 37 &   0.320541 	\\
6 & 5,000 &  2,500 & 29 &   0.320422 	\\
 \hline
\end{tabular}
\caption{Convergence of the estimation kernels with respect to the parameters for the discrete Fourier Transforms used to compute CSDs.}
\label{tabA1}
\end{table}
Constructing estimation kernels using a data-driven approach requires careful selection of discrete Fourier transform (DFT) parameters. Specifically, the streaming Fourier transform \citep{schmidt_efficient_2019} depends on several key parameters, including the number of frequencies ($N_{freq}$), the number of data blocks ($N_{blk}$), and the number of overlapping windows ($N_{ovlp}$). To reduce spectral leakage between blocks ($N_{blk}$), we apply a Hamming window with an overlap of $50\%$ \citep{trethewey_window_2000}. Table~\ref{tabA1} summarizes a convergence study evaluating how these parameters influence estimation accuracy within the spanwise-averaged flow under best sensor placement conditions. Based on this analysis, parameter set 6 was selected, and these DFT parameters were subsequently applied to the spanwise-Fourier and mid-span plane cases examined in this study.

\end{appendices}

\bibliographystyle{plainnat}
\bibliography{arXiv_template}

\begin{thebibliography}{58}
\providecommand{\natexlab}[1]{#1}
\providecommand{\url}[1]{\texttt{#1}}
\expandafter\ifx\csname urlstyle\endcsname\relax
  \providecommand{\doi}[1]{doi: #1}\else
  \providecommand{\doi}{doi: \begingroup \urlstyle{rm}\Url}\fi

\bibitem[Amaral and Cavalieri(2023)]{do_amaral_large-eddy-simulation-informed_2023}
F.~R. Amaral and A.~V.~G. Cavalieri.
\newblock Large-eddy-simulation-informed resolvent-based estimation of turbulent pipe flow.
\newblock \emph{Phys. Rev. Fluids}, 8\penalty0 (7):\penalty0 074606, July 2023.
\newblock ISSN 2469-990X.
\newblock \doi{10.1103/PhysRevFluids.8.074606}.

\bibitem[Amaral et~al.(2021)Amaral, Cavalieri, Martini, Jordan, and Towne]{amaral_resolvent-based_2021}
F.~R. Amaral, A.~V.~G. Cavalieri, E.~Martini, P.~Jordan, and A.~Towne.
\newblock Resolvent-based estimation of turbulent channel flow using wall measurements.
\newblock \emph{J. Fluid Mech.}, 927:\penalty0 A17, November 2021.

\bibitem[Audiffred et~al.(2023)Audiffred, Cavalieri, Brito, and Martini]{audiffred_experimental_2023}
D.~B.~S. Audiffred, A.~V.~G. Cavalieri, P.~P.~C. Brito, and E.~Martini.
\newblock Experimental control of {Tollmien}-{Schlichting} waves using the {Wiener}-{Hopf} formalism.
\newblock \emph{Phys. Rev. Fluids}, 8\penalty0 (7):\penalty0 073902, July 2023.

\bibitem[Audiffred et~al.(2024)Audiffred, Cavalieri, Maia, Martini, and Jordan]{audiffred_reactive_2024}
D.~B.~S. Audiffred, A.~V.~G. Cavalieri, I.~A. Maia, E.~Martini, and P.~Jordan.
\newblock Reactive experimental control of turbulent jets.
\newblock \emph{J. Fluid Mech.}, 994:\penalty0 A15, September 2024.
\newblock ISSN 0022-1120, 1469-7645.
\newblock \doi{10.1017/jfm.2024.569}.

\bibitem[Balay et~al.(2019)Balay, Abhyankar, Adams, and Others]{PETSC2019}
S.~Balay, S.~Abhyankar, M.~Adams, and Others.
\newblock Petsc users manual.
\newblock \emph{Argonne National Laboratory}, 2019.

\bibitem[Bhagwat(2021)]{rutvij}
R.~Bhagwat.
\newblock Development of stability analysis tools for high speed compressible flows.
\newblock \emph{PhD thesis}, 2021.

\bibitem[Br{\`e}s et~al.(2022)Br{\`e}s, Bose, Ivey, Emory, and Ham]{bres2022gpu}
G.~A. Br{\`e}s, S.~T. Bose, C.~B. Ivey, M.~Emory, and F.~Ham.
\newblock Gpu-accelerated large-eddy simulations of supersonic jets from twin rectangular nozzles.
\newblock \emph{AIAA Paper $\#$2022-3001}, 2022.

\bibitem[Brès et~al.(2017)Brès, Ham, Nichols, and Lele]{bres_unstructured_2017}
G.~A. Brès, F.~E. Ham, J.~W. Nichols, and S.~K. Lele.
\newblock Unstructured large-eddy simulations of supersonic jets.
\newblock \emph{AIAA J.}, 2017.

\bibitem[Brès et~al.(2018)Brès, Jordan, Jaunet, Le~Rallic, Cavalieri, Towne, Lele, Colonius, and Schmidt]{bres_importance_2018}
G.~A. Brès, P.~Jordan, V.~Jaunet, M.~Le~Rallic, A.~V.~G. Cavalieri, A.~Towne, S.~K. Lele, T.~Colonius, and O.~T. Schmidt.
\newblock Importance of the nozzle-exit boundary-layer state in subsonic turbulent jets.
\newblock \emph{J. Fluid Mech.}, 851:\penalty0 83--124, September 2018.
\newblock ISSN 0022-1120, 1469-7645.
\newblock \doi{10.1017/jfm.2018.476}.

\bibitem[Chevalier et~al.(2006)Chevalier, Hœpffner, Bewley, and Henningson]{chevalier_state_2006}
M.~Chevalier, J.~Hœpffner, T.~R. Bewley, and D.~S. Henningson.
\newblock State estimation in wall-bounded flow systems. {Part} 2. {Turbulent} flows.
\newblock \emph{J. Fluid Mech.}, 552\penalty0 (-1):\penalty0 167, March 2006.
\newblock ISSN 0022-1120, 1469-7645.
\newblock \doi{10.1017/S0022112005008578}.

\bibitem[Colonius and Towne(2025)]{duraisamy_chapter_2025}
T.~Colonius and A.~Towne.
\newblock Chapter 2 - {Modal} decomposition.
\newblock In Karthik Duraisamy, editor, \emph{Computation and {Analysis} of {Turbulent} {Flows}}, pages 27--81. Academic Press, 2025.
\newblock ISBN 978-0-323-95043-5.
\newblock \doi{https://doi.org/10.1016/B978-0-32-395043-5.00008-5}.

\bibitem[Daniele and Lombardi(2007)]{daniele_fredholm_2007}
V.~Daniele and G.~Lombardi.
\newblock Fredholm factorization of {Wiener}‐{Hopf} scalar and matrix kernels.
\newblock \emph{Radio Science}, 42\penalty0 (6):\penalty0 2007RS003673, December 2007.

\bibitem[Ducoin et~al.(2016)Ducoin, Loiseau, and Robinet]{ducoin_numerical_2016}
A.~Ducoin, J.-Ch. Loiseau, and J.-Ch. Robinet.
\newblock Numerical investigation of the interaction between laminar to turbulent transition and the wake of an airfoil.
\newblock \emph{European J. Mech. - B/Fluids}, 57:\penalty0 231--248, May 2016.
\newblock ISSN 09977546.
\newblock \doi{10.1016/j.euromechflu.2016.01.005}.

\bibitem[Farghadan et~al.(2024)Farghadan, Jung, Bhagwat, and Towne]{farghadan_efficient_2024}
A.~Farghadan, J.~Jung, R.~Bhagwat, and A.~Towne.
\newblock Efficient harmonic resolvent analysis via time stepping.
\newblock \emph{Theor. Comput. Fluid Dyn.}, May 2024.
\newblock ISSN 0935-4964, 1432-2250.
\newblock \doi{10.1007/s00162-024-00694-1}.

\bibitem[Farghadan et~al.(2025)Farghadan, Martini, and Towne]{farghadan_scalable_2025}
A.~Farghadan, E.~Martini, and A.~Towne.
\newblock Scalable resolvent analysis for three-dimensional flows.
\newblock \emph{J. Comput. Phys.}, 524:\penalty0 113695, March 2025.
\newblock ISSN 00219991.
\newblock \doi{10.1016/j.jcp.2024.113695}.

\bibitem[Fischer et~al.(2008)Fischer, Lottes, and Kerkemeier]{nek5000}
P.~F. Fischer, J.~W. Lottes, and S.~G. Kerkemeier.
\newblock Nek5000: Open source spectral element cfd solver, 2008.

\bibitem[Frigo and Johnson(2005)]{frigo_design_2005}
M.~Frigo and S.G. Johnson.
\newblock The {Design} and {Implementation} of {FFTW3}.
\newblock \emph{Proc. IEEE}, 93\penalty0 (2):\penalty0 216--231, February 2005.

\bibitem[Gartshore(1967)]{gartshore_two-dimensional_1967}
I.~S. Gartshore.
\newblock Two-dimensional turbulent wakes.
\newblock \emph{J. Fluid Mech.}, 30\penalty0 (3):\penalty0 547--560, November 1967.
\newblock ISSN 0022-1120, 1469-7645.
\newblock \doi{10.1017/S0022112067001600}.

\bibitem[Ghaemi and Scarano(2011)]{ghaemi_counter-hairpin_2011}
S.~Ghaemi and F.~Scarano.
\newblock Counter-hairpin vortices in the turbulent wake of a sharp trailing edge.
\newblock \emph{J. Fluid Mech.}, 689:\penalty0 317--356, December 2011.
\newblock ISSN 0022-1120, 1469-7645.
\newblock \doi{10.1017/jfm.2011.431}.

\bibitem[Gupta et~al.(2023)Gupta, Zhao, Sharma, Agrawal, Hourigan, and Thompson]{gupta_two-_2023}
S.~Gupta, J.~Zhao, A.~Sharma, A.~Agrawal, K.~Hourigan, and M.~C. Thompson.
\newblock Two- and three-dimensional wake transitions of a {NACA0012} airfoil.
\newblock \emph{J. Fluid Mech.}, 954:\penalty0 A26, January 2023.
\newblock ISSN 0022-1120, 1469-7645.
\newblock \doi{10.1017/jfm.2022.958}.

\bibitem[Hamming(1997)]{hamming1997digital}
R.~W. Hamming.
\newblock \emph{Digital filters}.
\newblock Dover Publications, 1997.

\bibitem[Hočevar et~al.(2005)Hočevar, Širok, and Grabec]{hocevar_turbulent-wake_2005}
M.~Hočevar, B.~Širok, and I.~Grabec.
\newblock A {Turbulent}-{Wake} {Estimation} {Using} {Radial} {Basis} {Function} {Neural} {Networks}.
\newblock \emph{Flow Turbul. Combust.}, 74\penalty0 (3):\penalty0 291--308, April 2005.
\newblock ISSN 1386-6184, 1573-1987.
\newblock \doi{10.1007/s10494-005-5728-4}.

\bibitem[Jin et~al.(2022)Jin, Illingworth, and Sandberg]{jin_optimal_2022}
B.~Jin, S.~J. Illingworth, and R.~D. Sandberg.
\newblock Optimal sensor and actuator placement for feedback control of vortex shedding.
\newblock \emph{J. Fluid Mech.}, 932:\penalty0 A2, February 2022.

\bibitem[Jovanović and Bamieh(2005)]{jovanovic_componentwise_2005}
M.~R. Jovanović and B.~Bamieh.
\newblock Componentwise energy amplification in channel flows.
\newblock \emph{J. Fluid Mech.}, 534:\penalty0 145--183, June 2005.

\bibitem[Jung(2024)]{Jung_RSV}
J.~Jung.
\newblock Resovlent-based estimation and control of aerodynamic flows.
\newblock \emph{PhD thesis}, 2024.

\bibitem[Jung and Towne(2024)]{jung_toward_2024}
J.~Jung and A.~Towne.
\newblock Toward turbulent wake estimation using a resolvent-based approach.
\newblock \emph{AIAA Paper $\#$2024-0057}, 2024.

\bibitem[Jung et~al.(2020)Jung, Martini, Cavalieri, Jordan, Lesshafft, and Towne]{Jung2020}
J.~Jung, E.~Martini, A.~Cavalieri, P.~Jordan, L.~Lesshafft, and A.~Towne.
\newblock Optimal resolvent-based estimation for flow control.
\newblock \emph{APS Division of Fluid Dynamics}, G06:\penalty0 004, 2020.

\bibitem[Jung et~al.(2023)Jung, Bhagwat, and Towne]{jung_resolvent-based_2023}
J.~Jung, R.~Bhagwat, and A.~Towne.
\newblock Resolvent-based estimation of laminar flow around an airfoil.
\newblock \emph{AIAA Paper $\#$2023-0077}, 2023.

\bibitem[Jung et~al.(2024)Jung, Bhagwat, and Towne]{jung_resolvent-based_2024}
J.~Jung, R.~Bhagwat, and A.~Towne.
\newblock Resolvent-based estimation and control of a laminar airfoil wake, December 2024.
\newblock arXiv:2412.19386.

\bibitem[Jung et~al.(2025)Jung, Bhagwat, and Towne]{jung_resovlent_based_2025_JFM}
J.~Jung, R.~Bhagwat, and A.~Towne.
\newblock Resolvent-based estimation and control of a laminar airfoil wake.
\newblock \emph{J. Fluid Mech.}, 0, 2025.
\newblock \doi{10.1017/jfm.2025.10423}.

\bibitem[Kalman(1960)]{Kalman_1960}
R.E. Kalman.
\newblock A new approach to linear filtering and prediction problems.
\newblock \emph{J. Basic Engng}, 82:\penalty0 35--45, 1960.

\bibitem[Kim et~al.(2009)Kim, Yang, Chang, and Chung]{kim2009}
D.-H. Kim, J.-H. Yang, J.-W. Chang, and J~Chung.
\newblock Boundary layer and near-wake measurements of naca 0012 airfoil at low reynolds numbers.
\newblock \emph{AIAA Paper $\#$2009-1472}, 2009.

\bibitem[Kojima et~al.(2013)Kojima, Nonomura, Oyama, and Fujii]{kojima_large-eddy_2013-1}
R.~Kojima, T.~Nonomura, A.~Oyama, and K.~Fujii.
\newblock Large-{Eddy} {Simulation} of {Low}-{Reynolds}-{Number} {Flow} {Over} {Thick} and {Thin} {NACA} {Airfoils}.
\newblock \emph{J. Aircraft}, 50\penalty0 (1):\penalty0 187--196, 2013.

\bibitem[Liu et~al.(2021)Liu, Sun, Yeh, Ukeiley, Cattafesta, and Taira]{liu_unsteady_2021}
Q.~Liu, Y.~Sun, C.~Yeh, L.~S. Ukeiley, L.~N. Cattafesta, and K.~Taira.
\newblock Unsteady control of supersonic turbulent cavity flow based on resolvent analysis.
\newblock \emph{J. Fluid Mech.}, 925:\penalty0 A5, October 2021.
\newblock ISSN 0022-1120, 1469-7645.

\bibitem[Maia et~al.(2021)Maia, Jordan, Cavalieri, Martini, Sasaki, and Silvestre]{maia_real-time_2021}
I.~A. Maia, P.~Jordan, A.~V.~G. Cavalieri, E.~Martini, K.~Sasaki, and F.~J. Silvestre.
\newblock Real-time reactive control of stochastic disturbances in forced turbulent jets.
\newblock \emph{Phys. Rev. Fluids}, 6\penalty0 (12):\penalty0 123901, December 2021.
\newblock ISSN 2469-990X.
\newblock \doi{10.1103/PhysRevFluids.6.123901}.

\bibitem[Manohar et~al.(2018)Manohar, Brunton, Kutz, and L.]{Manohar_data-driven_2018}
K.~Manohar, B.~W. Brunton, J.~N. Kutz, and Brunton~S. L.
\newblock Data-{Driven} {Sparse} {Sensor} {Placement} for {Reconstruction}: {Demonstrating} the {Benefits} of {Exploiting} {Known} {Patterns}.
\newblock \emph{IEEE Control Syst.}, 38\penalty0 (3):\penalty0 63--86, June 2018.
\newblock ISSN 1066-033X, 1941-000X.
\newblock \doi{10.1109/MCS.2018.2810460}.

\bibitem[Martinelli(2009)]{martinelli_feedback_nodate}
F.~Martinelli.
\newblock Feedback {Control} of {Turbulent} {Wall} {Flows}.
\newblock \emph{PhD thesis}, 2009.

\bibitem[Martini et~al.(2020)Martini, Cavalieri, Jordan, Towne, and Lesshafft]{martini_resolvent-based_2020}
E.~Martini, A.~V.~G. Cavalieri, P.~Jordan, A.~Towne, and L.~Lesshafft.
\newblock Resolvent-based optimal estimation of transitional and turbulent flows.
\newblock \emph{J. Fluid Mech.}, 900:\penalty0 A2, October 2020.
\newblock ISSN 0022-1120, 1469-7645.

\bibitem[Martini et~al.(2022)Martini, Jung, Cavalieri, Jordan, and Towne]{martini_resolvent-based_2022}
E.~Martini, J.~Jung, A.~Cavalieri, P.~Jordan, and A.~Towne.
\newblock Resolvent-based tools for optimal estimation and control via the {Wiener}–{Hopf} formalism.
\newblock \emph{J. Fluid Mech.}, 937:\penalty0 A19, April 2022.

\bibitem[McKeon and Sharma(2010)]{mckeon_critical-layer_2010}
B.~J. McKeon and A.~S. Sharma.
\newblock A critical-layer framework for turbulent pipe flow.
\newblock \emph{J. Fluid Mech.}, 658:\penalty0 336--382, September 2010.

\bibitem[Munday and Taira(2018)]{munday_effects_2018}
P.~M. Munday and K.~Taira.
\newblock Effects of wall-normal and angular momentum injections in airfoil separation control.
\newblock \emph{AIAA J.}, 2018.

\bibitem[Nielsen and Kleb(2006)]{nielsen_efficient_2006}
E.~J. Nielsen and W.~L. Kleb.
\newblock Efficient construction of discrete adjoint operators on unstructured grids using complex variables.
\newblock \emph{AIAA J.}, 2006.

\bibitem[Noble(1958)]{noble1958}
B.~Noble.
\newblock \emph{Methods Based on the Wiener-Hopf Technique for the Solution of Partial Differential Equations}, volume~7 of \emph{Int. Ser. Monogr. Pure Appl. Math.}
\newblock Pergamon Press, New York, 1958.

\bibitem[Reynolds(1972)]{reynolds_large-scale_1972}
W.~C. Reynolds.
\newblock Large-scale instabilities of turbulent wakes.
\newblock \emph{J. Fluid Mech.}, 54\penalty0 (3):\penalty0 481--488, August 1972.
\newblock ISSN 0022-1120, 1469-7645.
\newblock \doi{10.1017/S0022112072000813}.

\bibitem[Sashittal and Bodony(2021)]{Palash_DataDriven_2021}
P.~Sashittal and D.~J. Bodony.
\newblock Data-driven sensor placement for fluid flows.
\newblock \emph{AIAA Paper $\#$2021-2824}, 2021.

\bibitem[Schauerte and Schreyer(2024)]{schauerte_experimental_2024}
C.~J. Schauerte and A.~Schreyer.
\newblock Experimental investigation on the turbulent wake flow in fully established transonic buffet conditions.
\newblock \emph{CEAS Aeronautical J.}, 15\penalty0 (1):\penalty0 125--147, January 2024.
\newblock ISSN 1869-5582, 1869-5590.
\newblock \doi{10.1007/s13272-023-00690-x}.

\bibitem[Schmid and Sipp(2016)]{schmid_linear_2016}
P.~J. Schmid and D.~Sipp.
\newblock Linear control of oscillator and amplifier flows.
\newblock \emph{Phys. Fluids}, 1\penalty0 (4):\penalty0 040501, August 2016.
\newblock ISSN 2469-990X.

\bibitem[Schmidt and Towne(2019)]{schmidt_efficient_2019}
O.~T. Schmidt and A.~Towne.
\newblock An efficient streaming algorithm for spectral proper orthogonal decomposition.
\newblock \emph{Comput. Phys. Communi.}, 237:\penalty0 98--109, April 2019.
\newblock ISSN 00104655.
\newblock \doi{10.1016/j.cpc.2018.11.009}.

\bibitem[Shamsoddin and Porté-Agel(2017)]{shamsoddin_turbulent_2017}
S.~Shamsoddin and F.~Porté-Agel.
\newblock Turbulent planar wakes under pressure gradient conditions.
\newblock \emph{J. Fluid Mech.}, 830:\penalty0 R4, November 2017.
\newblock ISSN 0022-1120, 1469-7645.
\newblock \doi{10.1017/jfm.2017.649}.

\bibitem[Theofilis(2003)]{theofilis_advances_2003}
V.~Theofilis.
\newblock Advances in global linear instability analysis of nonparallel and three-dimensional flows.
\newblock \emph{Prog. in Aerosp. Sci.}, 39\penalty0 (4):\penalty0 249--315, May 2003.
\newblock ISSN 03760421.
\newblock \doi{10.1016/S0376-0421(02)00030-1}.

\bibitem[Towne et~al.(2018)Towne, Schmidt, and Colonius]{towne_spectral_2018}
A.~Towne, O.~T. Schmidt, and T.~Colonius.
\newblock Spectral proper orthogonal decomposition and its relationship to dynamic mode decomposition and resolvent analysis.
\newblock \emph{J. Fluid Mech.}, 847:\penalty0 821--867, July 2018.

\bibitem[Towne et~al.(2020)Towne, Lozano-Durán, and Yang]{towne_resolvent-based_2020}
A.~Towne, A.~Lozano-Durán, and X.~Yang.
\newblock Resolvent-based estimation of space–time flow statistics.
\newblock \emph{J. Fluid Mech.}, 883:\penalty0 A17, January 2020.
\newblock ISSN 0022-1120, 1469-7645.

\bibitem[Towne et~al.(2023)Towne, Dawson, Brès, Lozano-Durán, Saxton-Fox, Parthasarathy, Jones, Biler, Yeh, Patel, and Taira]{towne_database_2023}
A.~Towne, S.~T.~M. Dawson, G.~A. Brès, A.~Lozano-Durán, T.~Saxton-Fox, A.~Parthasarathy, A.~R. Jones, H.~Biler, C.~Yeh, H.~D. Patel, and K.~Taira.
\newblock A database for reduced-complexity modeling of fluid flows.
\newblock \emph{AIAA J.}, 2023.

\bibitem[Towne et~al.(2024)Towne, Bhagwat, Zhou, Jung, Martini, Jordan, Audiffred, Maia, and Cavalieri]{towne_resolvent-based_2024}
A.~Towne, R.~Bhagwat, Y.~Zhou, J.~Jung, E.~Martini, P.~Jordan, D.~Audiffred, I.~Maia, and A.~Cavalieri.
\newblock Resolvent-based estimation of wavepackets in turbulent jets.
\newblock \emph{AIAA Paper $\#$2024-3413}, 2024.

\bibitem[Trethewey(2000)]{trethewey_window_2000}
M.W. Trethewey.
\newblock Window and overlap processing effects on power estimates from spectra.
\newblock \emph{Mech. Syst. Signal Process.}, 14\penalty0 (2):\penalty0 267--278, March 2000.
\newblock ISSN 0888-3270.
\newblock \doi{10.1006/mssp.1999.1274}.
\newblock Publisher: Elsevier BV.

\bibitem[Yeh and Taira(2019)]{yeh_resolvent-analysis-based_2019}
C.~Yeh and K.~Taira.
\newblock Resolvent-analysis-based design of airfoil separation control.
\newblock \emph{J. Fluid Mech.}, 867:\penalty0 572--610, May 2019.
\newblock ISSN 0022-1120, 1469-7645.

\bibitem[Yeh et~al.(2020)Yeh, Benton, Taira, and Garmann]{yeh_resolvent_2020}
C.~Yeh, S.~I. Benton, K.~Taira, and D.~J. Garmann.
\newblock Resolvent analysis of an airfoil laminar separation bubble at {Re} = 500 000.
\newblock \emph{Phys. Rev. Fluids}, 5\penalty0 (8):\penalty0 083906, August 2020.
\newblock ISSN 2469-990X.

\bibitem[Ying et~al.(2023)Ying, Liang, Li, and Fu]{ying_resolvent-based_2023}
A.~Ying, T.~Liang, Z.~Li, and L.~Fu.
\newblock A resolvent-based prediction framework for incompressible turbulent channel flow with limited measurements.
\newblock \emph{J. Fluid Mech.}, 976:\penalty0 A31, December 2023.
\newblock ISSN 0022-1120, 1469-7645.
\newblock \doi{10.1017/jfm.2023.867}.

\end{thebibliography}

\end{document}